\documentstyle[epsf,12pt]{article}

\makeatletter

 \@addtoreset{equation}{section}
\makeatother

\begin{document}
\topmargin 0pt
\oddsidemargin 5mm

\newcommand {\beq}{\begin{eqnarray}}
\newcommand {\eeq}{\end{eqnarray}}
\newcommand {\non}{\nonumber\\}
\newcommand {\eq}[1]{\label {eq.#1}}
\newcommand {\defeq}{\stackrel{\rm def}{=}}
\newcommand {\gto}{\stackrel{g}{\to}}
\newcommand {\hto}{\stackrel{h}{\to}}
\newcommand {\1}[1]{\frac{1}{#1}}
\newcommand {\2}[1]{\frac{i}{#1}}
\newcommand {\th}{\theta}
\newcommand {\thb}{\bar{\theta}}
\newcommand {\ps}{\psi}
\newcommand {\psb}{\bar{\psi}}
\newcommand {\ph}{\varphi}
\newcommand {\phs}[1]{\varphi^{*#1}}
\newcommand {\sig}{\sigma}
\newcommand {\sigb}{\bar{\sigma}}
\newcommand {\Ph}{\Phi}
\newcommand {\Phd}{\Phi^{\dagger}}
\newcommand {\Sig}{\Sigma}
\newcommand {\Phm}{{\mit\Phi}}
\newcommand {\eps}{\varepsilon}
\newcommand {\del}{\partial}
\newcommand {\dagg}{^{\dagger}}
\newcommand {\pri}{^{\prime}}
\newcommand {\prip}{^{\prime\prime}}
\newcommand {\pripp}{^{\prime\prime\prime}}
\newcommand {\prippp}{^{\prime\prime\prime\prime}}
\newcommand {\pripppp}{^{\prime\prime\prime\prime\prime}}
\newcommand {\delb}{\bar{\partial}}
\newcommand {\zb}{\bar{z}}
\newcommand {\mub}{\bar{\mu}}
\newcommand {\nub}{\bar{\nu}}
\newcommand {\lam}{\lambda}
\newcommand {\lamb}{\bar{\lambda}}
\newcommand {\kap}{\kappa}
\newcommand {\kapb}{\bar{\kappa}}
\newcommand {\xib}{\bar{\xi}}
\newcommand {\ep}{\epsilon}
\newcommand {\epb}{\bar{\epsilon}}
\newcommand {\Ga}{\Gamma}
\newcommand {\rhob}{\bar{\rho}}
\newcommand {\etab}{\bar{\eta}}
\newcommand {\chib}{\bar{\chi}}
\newcommand {\tht}{\tilde{\th}}
\newcommand {\zbasis}[1]{\del/\del z^{#1}}
\newcommand {\zbbasis}[1]{\del/\del \bar{z}^{#1}}
\newcommand {\vecv}{\vec{v}^{\, \prime}}
\newcommand {\vecvd}{\vec{v}^{\, \prime \dagger}}
\newcommand {\vecvs}{\vec{v}^{\, \prime *}}
\newcommand {\alpht}{\tilde{\alpha}}
\newcommand {\xipd}{\xi^{\prime\dagger}}
\newcommand {\pris}{^{\prime *}}
\newcommand {\prid}{^{\prime \dagger}}
\newcommand {\Jto}{\stackrel{J}{\to}}
\newcommand {\vprid}{v^{\prime 2}}
\newcommand {\vpriq}{v^{\prime 4}}
\newcommand {\vt}{\tilde{v}}
\newcommand {\vecvt}{\vec{\tilde{v}}}
\newcommand {\vecpht}{\vec{\tilde{\phi}}}
\newcommand {\pht}{\tilde{\phi}}
\newcommand {\goto}{\stackrel{g_0}{\to}}
\newcommand {\tr}{{\rm tr}\,}
\newcommand {\GC}{G^{\bf C}}
\newcommand {\HC}{H^{\bf C}}
\newcommand{\vs}[1]{\vspace{#1 mm}}
\newcommand{\hs}[1]{\hspace{#1 mm}}

\setcounter{page}{0}

\begin{titlepage}

\begin{flushright}
OU-HET 332\\
hep-th/9911139\\
November 1999
\end{flushright}
\bigskip

\begin{center}
{\LARGE\bf
Supersymmetric Nonlinear Sigma Models as Gauge Theories
}
\vs{10}

\bigskip
{\renewcommand{\thefootnote}{\fnsymbol{footnote}}
{\large\bf Kiyoshi Higashijima\footnote{
     e-mail: higashij@phys.sci.osaka-u.ac.jp.}
 and Muneto Nitta\footnote{
e-mail: nitta@het.phys.sci.osaka-u.ac.jp.}
}}

\setcounter{footnote}{0}
\bigskip

{\small \it
Department of Physics,
Graduate School of Science, Osaka University,\\
Toyonaka, Osaka 560-0043, Japan
}
\end{center}
\bigskip

\begin{abstract}
Supersymmetric nonlinear sigma models are obtained from 
linear sigma models by imposing supersymmetric constraints. 
If we introduce auxiliary chiral and vector superfields,  
these constraints can be expressed by D-terms and F-terms 
depending on the target manifolds. Auxiliary vector 
superfields appear as gauge fields without kinetic terms.
If there are no D-term constraints, the target manifolds 
are always non-compact manifolds. When all the degrees 
of freedom in these non-compact directions are eliminated 
by gauge symmetries, the target manifold becomes compact. 
All supersymmetric nonlinear sigma models, whose target 
manifolds are the hermitian symmetric spaces, are 
successfully formulated as gauge theories.
\end{abstract}

\end{titlepage}

\section{Introduction}

When the global symmetry $G$ is spontaneously broken down to its 
subgroup $H$, there appear massless Nambu-Goldstone (NG) bosons 
corresponding to broken generators of the coset manifold $G/H$. 
At low energies, interactions among these massless particles are 
described by the so-called nonlinear sigma models, whose 
lagrangians are completely determined by the geometry of the 
target manifold $G/H$, parameterized by NG-bosons~\cite{CCWZ}.

In supersymmetric theories, there appear massless fermions 
as supersymmetric partners of NG-bosons~\cite{BPY}. These 
massless fermions together with NG-bosons are described by 
chiral superfields in four dimensional theories with $N=1$ 
supersymmetry. Since chiral superfields are complex, the 
supersymmetric nonlinear sigma models are closely related to 
the complex geometry; their target manifolds, where fields 
variables take their values, must be K\"{a}hler 
manifolds~\cite{Zu}. 
If the coset manifold $G/H$ itself happens to be a K\"{a}hler 
manifold, both real and imaginary parts of the scalar 
components of chiral superfields are NG-bosons. If $G/H$ 
is not a K\"{a}hler manifold, on the other hand, there is 
at least one chiral superfield whose real or imaginary part 
is not a NG-boson. This additional massless boson is called 
the quasi-Nambu-Goldstone (QNG) boson~\cite{BPY,KOY}. 

The general method to construct supersymmetric nonlinear sigma 
models has been discussed by Bando, Kuramoto, Maskawa and Uehara
(BKMU)~\cite{BKMU}. 
When QNG bosons are present, their effective 
lagrangians include arbitrary functions. This is always the case 
when the target manifold of the nonlinear sigma model is larger 
than the coset manifold $G/H$, where NG-bosons reside, since the 
geometry of the target manifold cannot be fixed by the metric of 
its subspace $G/H$~\cite{LS}--\cite{Ni2}. 
The arbitrariness reflects the ambiguity of 
the metric in the direction of QNG bosons. When the coset 
manifold $G/H$ is itself K\"{a}hler, the effective lagrangian is 
uniquely determined by the geometry of $G/H$, as has been shown 
in a beautiful paper by Itoh, Kugo and Kunitomo~\cite{IKK}.
(See Appendix A for a review.) 
K\"{a}hler potentials in this case 
have been discussed by many authors~\cite{BKMU,IKK}--\cite{BKY}
(See references in Ref.~\cite{BKY}.), 
and have been used to construct the coset unification models, 
where fermionic partners of NG bosons are 
considered as quarks~\cite{CUM}.

Nonlinear sigma models are considered low energy 
effective theories for massless particles after integrating out 
the massive particles in the corresponding linear sigma models. 
In this context, Lerche and Shore have shown that nonlinear sigma 
models whose target manifolds are K\"{a}hler $G/H$ manifolds 
cannot be obtained from linear sigma models~\cite{LS}. 
(See also Ref.~\cite{non-compact} and Appendix B for a review.) 
According to this theorem, there must exist at least one 
QNG bosons in effective field theories obtained from 
linear sigma models. 

On the other hand, it is known that sigma models on some 
K\"{a}hler $G/H$ manifolds, namely on ${\bf C}P^N$ or on 
the Grassmann manifold $G_{N,M}({\bf C})$, are obtained by 
the introduction of gauge symmetry~\cite{Ao,LR,HKLR,Ku}. 
The implicit assumption of Lerche and Shore is the absence of gauge 
interactions in the linear sigma models. It seems possible 
to eliminate unnecessary QNG bosons if we introduce 
an appropriate gauge symmetry. 

In this paper, we show that supersymmetric nonlinear sigma 
models on a certain class of K\"{a}hler $G/H$ manifolds 
are obtained from linear sigma models with gauge symmetry. 
We define nonlinear sigma models by imposing 
supersymmetric constraints on linear sigma models. 
We introduce two kinds of constraints, D-term and F-term 
constraints. If we introduce auxiliary fields, these 
correspond to vector and chiral superfields. Vector 
auxiliary superfields appear as gauge fields. 
We successfully formulate nonlinear sigma models on 
(irreducible and compact) hermitian symmetric spaces\footnote{ 
Symmetric spaces are homogeneous spaces $G/H$ 
with an involutive automorphism. 
Since it can be shown that 
any $G$-invariant differential form $\omega$  
in a symmetric space is closed, $d\omega =0$, 
a fundamental two form of a hermitian symmetric space 
is also closed, and this is K\"{a}hlerian.
Hence, the expression ``K\"{a}hler symmetric space'' 
has the same meaning.}
classified by Cartan as in Table~1~\cite{CV}.\footnote{
We use `$\dim_{\bf C}$' for complex dimensions and `dim' 
for real dimensions.
}
\begin{table}
\caption{\bf Hermitian Symmetric Spaces}
\begin{center}
\begin{tabular}{|c|c|c|}
 \noalign{\hrule height0.8pt}
  Type & $G/H$ & $\dim_{\bf C} (G/H)$\\
 \hline
 \noalign{\hrule height0.2pt}
 AIII$_1$&${\bf C}P^{N-1}=SU(N)/S(U(N-1)\times U(1))$&$N-1$\\
 AIII$_2$&$G_{N,M}({\bf C})=U(N)/U(N-M)\times U(M)$  &$M(N-M)$\\
 BDI     &$Q^{N-2}({\bf C})=SO(N)/SO(N-2)\times U(1)$&$N-2$\\
 CI      &$Sp(N)/U(N)$                          &$\1{2}N(N+1)$\\
 DIII    &$SO(2N)/U(N)$                         &$\1{2}N(N-1)$\\     
 EIII    &$E_6/SO(10)\times U(1)$                    &$16$\\
 EVII    &$E_7/E_6 \times U(1)$                      &$27$\\   
 \noalign{\hrule height0.8pt}
 \end{tabular}
 \end{center}
\begin{footnotesize}
The first three manifolds, 
${\bf C}P^{N-1}$, $G_{N,M}({\bf C})$ and $Q^{N-2}({\bf C})$, 
are called a projective space, a Grassmann manifold 
and a quadratic surface, respectively. 
The projective space ${\bf C}P^{N-1}$ and 
the Grassmann manifold $G_{N,M}({\bf C})$
are a set of complex lines and $M$ dimensional complex planes  
in ${\bf C}^N$, respectively. 
BI (DI) corresponds to odd (even) $N$.
In the mathematical literature, 
EIII is written as $E_6/Spin(10)\times U(1)$, 
since coset generators belong to the $SO(10)$ Weyl spinor.
\end{footnotesize}
\end{table}

\bigskip
This paper is organized as follows. 
In Sec.~2, we review simple cases without F-term 
constraints, namely the projective space ${\bf C}P^{N-1}$ 
and the Grassmann manifold $G_{N,M}({\bf C})$. 
Although these cases are known, it is instructive to 
discuss them with emphasis on an interpretation 
in terms of NG and QNG bosons. 
In Sec.~3, we generalize to other hermitian symmetric spaces 
by introducing F-term constraints in addition 
to D-term constraints. Results in this section are new.
As a by-product, we find explicit expressions of 
holomorphic constraints to embed $G/H$ into 
${\bf C}P^N$ or $G_{N,M}({\bf C})$. 
Sec.~4 is devoted to conclusions and discussion.
We discuss how the results can be generalized to 
an arbitrary K\"{a}hler $G/H$ manifold. 
In Appendix A, we review the construction of the 
K\"{a}hler potentials for K\"{a}hler $G/H$ 
using BKMU and IKK methods, in the case of 
hermitian symmetric spaces. 
In Appendix B, we review the theorem of Lerche and Shore. 
Appendices C, D and E are devoted to 
summaries of $SO$, $E_6$ and $E_7$ algebras. 

\bigskip
In the rest of this section, we introduce the notation 
and terminology used in this paper. 

The linear description of the nonlinear sigma model 
without a gauge symmetry is given by 
\beq
 {\cal L} = \int d^4\th \phi\dagg \phi 
    + \left(\int d^2\th \phi_0 g (\phi)+ {\rm c.c.} \right), 
\eeq
where the chiral superfield $\phi$ belongs to an irreducible 
representation of the global symmetry group $G$, 
and $\phi_0$ is an auxiliary chiral superfield.
The absence of kinetic term of $\phi_0$ corresponds to the strong 
coupling limit of the Yukawa theory.
Although the superpotential $W = \phi_0 g(\phi)$ is 
$G$-invariant, $\phi_0$ and $g(\phi)$ need not be 
$G$-invariant separately. Instead, they may have indices 
transforming as a non-trivial representation of $G$, such as 
$W={\phi_0}_i g(\phi)^i$. 
If we integrate over the auxiliary field $\phi_0$, 
we obtain F-term constraints, $g(\phi)=0$, which are 
holomorphic functions. Therefore, the F-term 
constraints are invariant under the larger group $\GC$, 
the complex extension of $G$ . 
 
Let the number of F-term constraints be $N_{\rm F}$.
If it is sufficiently large, the target manifold $M\pri$ 
becomes a $\GC$-orbit of the vacuum $v=\left<\phi\right>$.  
Let the complex isotropy group of the vacuum be $\hat H$
\ ($\hat H v=v$).
Then, the target manifold of the nonlinear sigma model 
is parameterized by the chiral superfields corresponding 
to complex broken generators in 
${\cal \GC} -\hat{\cal H}$.\footnote{
We use the calligraphic font for a Lie algebra 
corresponding to a Lie group.} 
Therefore $M\pri$ is a complex coset space, 
$M\pri \simeq \GC/\hat H$, generated by these broken generators. 
As an example, let us consider a doublet 
$\phi=\pmatrix{ \phi_1 \cr \phi_2}$ of $G=SU(2)$ and suppose 
that they acquire the vacuum expectation values 
$v=\pmatrix{1 \cr 0}$.
Since the raising operator 
$\tau_+= \1{2} (\tau_1+i\tau_2) = \pmatrix{0 & 1 \cr 0 & 0}$ 
satisfies $\tau_+v=0$, it is the complex unbroken 
generator in $\hat{\cal H}$. On the other hand, $\tau_3$ 
and the lowering operator $\tau_- (= {\tau_+}\dagg)$ 
are the elements of 
the broken generators in ${\cal \GC} -\hat{\cal H}$.

There are two kinds of broken generators: the hermitian broken 
generator $X$ and 
the non-hermitian broken generator $E$.\footnote{
In general, $\hat H$ is larger than $H^{\bf C}$, 
due to the existence of non-hermitian generators $\bar E$. 
$\bar E$ is the hermitian conjugate of $E$.
They constitute the so-called Borel subalgebra ${\cal B}$ 
in ${\hat{\cal H}}$~\cite{BKMU}. 
}  
The superfields corresponding to non-hermitian 
and hermitian generators are called pure-type and 
mixed-type superfields, respectively~\cite{BKMU,LS}.
In the previous example, where the representative of 
$\GC/\hat H$ is given by 
$\phi =\exp{i(\varphi_3\tau_3+\varphi\tau_-)}\cdot v$, 
$\varphi_3$ is a mixed-type and $\varphi$ is 
a pure-type superfield. 
The scalar components of the mixed-type multiplets 
consist of a QNG boson in addition to a NG boson, 
whereas those of the pure-type multiplets 
consist of two genuine NG bosons. 
Since the vacuum is invariant under ${\hat H}$, 
we can multiply the representative of the coset manifold
by an arbitrary element of ${\hat H}$ from the right. 
In our previous example, we can rewrite it as 
$\exp{i(\varphi_3\tau_3+\Re\varphi\tau_1+\Im\varphi\tau_2)}
\cdot v$ by multiplying an appropriate factor generated by 
$\tau_+$ for sufficiently small $|\varphi_3|$ and 
$|\varphi|$. The NG-bosons parameterizing $S^3\simeq SU(2)$ 
are $\Re\varphi_3, \ \Re\varphi, \ \Im\varphi\ $, 
whereas $\Im\varphi_3$ is the QNG-boson parameterizing the 
radius of $S^3$. 
The number of chiral superfields parameterizing 
the target manifold is 
\beq
 N_{\Ph} 
 = \dim_{\bf C} V - N_{\rm F}
 = N_{\rm M} + N_{\rm P},
\eeq
where $V$ is the representation space. 
The numbers of the mixed-type and pure-type multiplets 
are denoted by $N_{\rm M}$ and $N_{\rm P}$, respectively. 

The directions parameterized by QNG bosons are non-compact, 
whereas those of NG bosons are compact.\footnote{
We use the word ``compactness'' in the sense of topology.
The kinetic terms of QNG bosons have the same sign as 
those of NG bosons.
}
From the theorem 
of Lerche and Shore (see Appendix B), 
there exists at least one mixed-type multiplet, 
and therefore the target manifold $M\pri$ 
becomes non-compact. Since no two points 
in the non-compact direction can be connected 
by the compact isometry group $G$, $M\pri$ is 
also non-homogeneous.

We rewrite the groups $G$ and $H$ 
defined above as $G\pri$ and $H\pri$, 
and therefore $M\pri \simeq G^{\prime {\bf C}}/\hat H\pri$.
In order to eliminate the degree of freedom of QNG bosons, 
we elevate the subgroup of $G\pri$ to a local gauge symmetry.
We assume $G\pri$ is the direct product of a global symmetry 
and the gauge symmetry $G_{\rm gauge}$; 
that is $G\pri=G\times G_{\rm gauge}$, 
where $G_{\rm gauge} = U(1)$ or $U(N)$. 
The gauged linear lagrangian can be written as 
\beq
 {\cal L} 
  = \int d^4\th 
    \left( e^V \phi\dagg\phi - c V\right) 
  + \left(\int d^2\th \phi_0 g(\phi)
          + {\rm c.c.} \right), 
\eeq
where $\phi_0$ and $V$ are auxiliary 
chiral and vector superfields. 
The absence of the kinetic term of the gauge field 
corresponds to the strong coupling limit, 
where the gauge coupling constant tends to infinity.
Here, for simplicity, the gauge group is 
assumed to be $U(1)$. 
(See Sec.~2.2 for the non-Abelian case.) 
Integration over $\phi_0$ gives the F-term constraint to define 
the non-compact manifold $M\pri$, as discussed above. 
The integration over $V$ gives a D-term constraint 
that restricts $M\pri$ to the compact manifold 
$M = M\pri/G_{\rm gauge}^{\bf C}$~\cite{HKLR}, 
whose dimension is 
\beq
 \dim_{\bf C} M = N_{\Phi} - \dim G_{\rm gauge} .
\eeq
Since we introduce gauge fields 
to absorb all mixed type multiplets,\footnote{
The supersymmetric Higgs mechanism acts as follows: 
A vector superfield absorbs one mixed-type multiplet 
to constitute a massive vector multiplet.
If it absorbs a pure-type multiplet, 
one NG boson remains massless. 
They cannot constitute a massive vector multiplet, 
and the supersymemtry is spontaneously broken~\cite{WB}.
} 
the dimension of the gauge group and the compact 
manifold $M$ are 
\beq
 \dim G_{\rm gauge} = N_{\rm M}, \hspace{1cm}
 \dim_{\rm C} M = N_{\rm P} .
\eeq
The compact manifold $M$ is parameterized by only 
pure-type multiplets. 

\section{Nonlinear Sigma Models without the F-term Constraint}
Although examples in this section are 
well known~\cite{HKLR}, 
we describe them in detail, since the interpretation in terms of NG and 
QNG bosons is useful to find the nonlinear sigma models on other 
compact manifolds.
 
\subsection{Projective space: 
${\bf C}P^{N-1}=SU(N)/S(U(N-1)\times U(1))$}
We consider the global symmetry 
$G\pri = U(N) = SU(N) \times U(1)_{\rm D} 
\defeq G \times U(1)_{\rm D}$. 
Below, the phase symmetry $U(1)_{\rm D}$ is gauged,  
while $G=SU(N)$ remains global.
We prepare the fundamental fields $\vec{\phi} \in {\bf N}$, 
which acquire a vacuum expectation value.
First of all, we consider the canonical K\"{a}hler potential 
\beq
 K(\vec{\phi},\vec{\phi}^{\,\dagger}) 
 = \vec{\phi}^{\,\dagger} \vec{\phi} .
\eeq
For later purposes, 
we decompose $G=SU(N)$ under the subgroup $SU(N-1) \times U(1)$.  
A fundamental representation ${\bf N}$ is decomposed as 
${\bf N}= ({\bf N-1},1)\oplus ({\bf 1},-N+1)$, 
where the second factors are $U(1)$ charges.
Hence, we decompose the fields as 
$\vec{\phi} = \pmatrix{ x \cr y^i}$ ($i=1,\cdots,N-1$).
Generators of $SU(N)$ can also be decomposed into
the $SU(N-1)$ generators $T_A$ $(A=1,\cdots,N^2-2N$), 
the $U(1)$ generator $T$, 
the $N-1$ raising operators $E^i$ represented 
by upper triangle matrices, 
and the lowering operators represented by 
lower triangle matrices $\bar E_i = (E^i)\dagg$.
The transformation law of $\vec{\phi}$  
under the complexified group $SU(N)^{\bf C}$ is
\beq
 \delta \vec{\phi}
 &=& \left(i \th T + i\th^A T_A 
    + \epb_i E^i + \ep^i \bar{E}_i \right) \vec{\phi} \non
 &=& \pmatrix 
 { i \sqrt{2(N-1) \over N}\th & \epb_j \cr  
   \ep^i & - i\th^A {\rho(T_A)^i}_j 
           - i \sqrt{2 \over N(N-1)} \th {\delta^i}_j\cr }
 \pmatrix{x \cr y^j} , \label{CPN-tr.law}
\eeq
where $\rho (T_A)$ is an $N-1$ by $N-1$ matrix for 
the fundamental representation of $SU(N-1)$.
We normalized these generators as 
$\tr {T_A}^2 = \tr T^2 = \tr \bar E_i E^i = 2$ (no sum).
When $\ep = \epb$ and $\th, \th^A \in {\bf R}$, 
this transformation law reduces to that of the real group $SU(N)$.
The $U(1)_{\rm D}$ transformation is generated by 
$T_{\rm D}={\bf 1}_N$.

When $\vec{\phi}$ acquires a vacuum expectation value, 
it can be transformed by ${G\pri}^{\bf C}$ 
to the standard form,
\beq
 \vec{v} = \left<\vec{\phi}\right> 
  = \pmatrix{1 \cr {\bf 0}} .\label{CPN-vev}
\eeq
By this vacuum, 
the global symmetry is spontaneously broken down as
$U(N) \to U(N-1) = SU(N-1) \times U(1)\pri \defeq H\pri$.
Here, $U(1)\pri$ is generated by 
$T\pri \sim {\rm diag}(0,1,\cdots,1)$,
which is a linear combination of $T_{\rm D}$ and $T$.  
The complex isotropy group $\hat H\pri$, which leaves $\vec{v}$ 
invariant, is larger than ${H\pri}^{\bf C}$, 
since upper triangle 
generators $E^i$ annihilate the vacuum $\vec{v}$. 
Here, $E^i$ generators constitute 
a Borel subalgebra ${\cal B}$ in $\hat {\cal H}\pri$.
On the other hand, the complex broken generators are 
the lower triangle generators $\bar E_i$ and 
the diagonal generator $X = (1,0,\cdots,0)$, 
which is also a linear combination of $T$ and $T_{\rm D}$. 
The non-hermitian generators $\bar E_i$ are pure-type generators,  
and the hermitian generator $X$ is a mixed-type generator.
The target manifold $M\pri$ of the nonlinear sigma model 
is a complex coset manifold 
$M\pri \simeq {G\pri}^{\bf C} /\hat H\pri$ 
generated by these complex broken generators. 
Since, by using its representative 
$\xi\pri = \exp( \ph^i \bar{E}_i + i\psi X)$, 
the fields can be written as $\vec{\phi} = \xi\pri \vec{v}$,  
its form near the vacuum is 
\beq
 \delta \vec{v} 
  = (i \psi X + \ph^i \bar{E}_i) \vec{v} 
  = \pmatrix{i \psi \cr \ph^i} .\label{CPN-MP}
\eeq
We thus find that $\psi$ is 
a mixed-type chiral superfield, 
whose scalar components are NG and QNG bosons, 
while the $\ph^i$ are pure-type chiral superfields, 
whose scalar components are both NG bosons.  
Then the numbers of mixed-type and 
pure-type chiral superfields are 
$N_{\rm M} = 1$ and $N_{\rm P} = N-1$, respectively.
This K\"{a}hler manifold is non-compact and non-homogeneous 
due to the existence of the QNG boson.

\bigskip
To construct a compact homogeneous manifold, 
we wish to eliminate the QNG boson (the mixed-type multiplet). 
Hence, we gauge the $U(1)_{\rm D}$ symmetry 
by introducing a vector superfield $V$, 
which will absorb the mixed-type multiplet. 
The gauged K\"{a}hler potential is~\cite{WB} 
\beq
 K(\vec{\phi},\vec{\phi}^{\,\dagger},V) 
 = e^V \vec{\phi}^{\,\dagger} \vec{\phi} -cV, \label{gauged-Kahler}
\eeq
where $c V$ is a Fayet-Iliopoulos (FI) D-term~\cite{HKLR,WB}. 
Since the transformation law of $V$ is
\beq
 e^V \to e^{V\pri} 
 = e^V e^{i(\th\dagg - \th)} ,\hspace{1cm}\; 
 e^{i \Re \th} \in U(1)_{\rm D},
\eeq
where $\th$ is a chiral superfield, 
the K\"{a}hler potential (\ref{gauged-Kahler}) is 
invariant under the complex extension of the gauge symmetry, 
${U(1)_{\rm D}}^{\bf C}$. 
Note that the global symmetry $G=SU(N)$ cannot be complexified. 
The equation of motion of $V$ is 
\beq
 \delta K/\delta V =  e^V \vec{\phi}^{\,\dagger}\vec{\phi} -c = 0.
\eeq
From this equation, $V$ can be solved as
\beq
 V(\vec{\phi},{\vec{\phi}}^{\,\dagger}) 
 = - \log \left({\vec{\phi}^{\,\dagger}\vec{\phi}\over c}\right) .
\eeq
To eliminate the gauge field, 
we substitute $V(\vec{\phi},{\vec{\phi}}^{\,\dagger})$ 
back into Eq.~(\ref{gauged-Kahler}), obtaining 
\beq
 K(\vec{\phi},\vec{\phi}^{\,\dagger},
   V(\vec{\phi},\vec{\phi}^{\,\dagger})) 
 = c \log (\vec{\phi}^{\,\dagger}\vec{\phi}),  \label{Kahler}
\eeq
where we have omitted constant terms.\footnote{
Their contributions to the lagrangian vanish as a result of 
the $d^4\theta$ integration.} 
Since we have the gauge symmetry ${U(1)_{\rm D}}^{\bf C}$,  
we can fix the gauge as 
\beq
 \vec{\phi} = \pmatrix{1 \cr \ph^i} .\label{CPN-phi}
\eeq
By comparing Eqs.~(\ref{CPN-MP}) and (\ref{CPN-phi}), 
we find that the mixed-type chiral superfield 
has been eliminated by this gauge fixing.
The gauge fixed field (\ref{CPN-phi}) can be rewritten as 
\beq
 \vec{\phi} = \xi \vec{v},\hspace{1cm} 
 \xi = e^{\ph \cdot \bar E} 
 = \pmatrix{1  & {\bf 0}\cr
         \ph^i & {\bf 1}_{N-1}}, \label{CPN-phi2}
\eeq
where $\xi$ can be considered as a representative of 
a complex coset manifold 
$G^{\bf C}/\hat H \simeq G/H = SU(N)/S(U(N-1)\times U(1))$. 
Since this is a compact homogeneous K\"{a}hler manifold, 
we have obtained the desired result.  
To obtain a compact manifold, gauge fields are necessary. 
By substituting Eq.~(\ref{CPN-phi}) into Eq.~(\ref{Kahler}), 
we obtain
\beq
 K(\ph,\ph\dagg,V(\ph,\ph\dagg)) = c \log (1 + |\ph|^2) . 
\eeq
This is the well-known K\"{a}hler potential of 
the Fubini-Study metric for 
${\bf C}P^{N-1} = SU(N)/S(U(N-1)\times U(1))$.

\subsection{Grassmann manifold: 
$G_{N,M}({\bf C}) =U(N)/U(N-M)\times U(M)$}
This subsection is a generalization of 
the last subsection. 
The picture of NG and QNG bosons 
is discussed in Ref.~\cite{Ku}.
We consider a global symmetry 
$G\pri = G_{\rm L} \times G_{\rm R} 
= U(N)_{\rm L} \times U(M)_{\rm R}\, (N>M)$. 
The basic fields are $\Phi \in ({\bf N},{\bf \bar M})$, 
which are $N \times M$ matrix-valued chiral superfields. 
The transformation law of $\Phi$ 
under ${G\pri}^{\bf C}$ is\footnote{
The conjugate representation $\vec{\phi} \in {\bf \bar N}$ 
is defined to transform as $\vec{\phi} \to (g^{-1})^T \vec{\phi}$, 
since the group is extended to its complexification and 
we must preserve the chirality.} 
\beq
 \Phi \to \Phi\pri 
 = g \cdot \Phi 
 \defeq g_{\rm L} \Phi {g_{\rm R}}^{-1}, \hspace{1cm}
 g =(g_{\rm L},g_{\rm R}) \in {G\pri}^{\bf C},
\eeq
where $g_{\rm L}$ and $g_{\rm R}$ are $N\times N$ and 
$M\times M$ matrices, respectively.

The K\"{a}hler potential is canonical:
\beq
 K (\Ph,\Ph\dagg) = \tr (\Ph\dagg\Ph) .
\eeq
Any vacuum can be transformed under ${G\pri}^{\bf C}$ to
\beq
 V = \left< \Ph \right> 
  = \pmatrix {{\bf 1}_M \cr
              {\bf 0}} ,
\eeq
where ${\bf 1}_M$ is the $M \times M$ identity matrix and 
$\bf 0$ is the $(N-M) \times M$ zero matrix.
The global symmetry is spontaneously broken as 
$U(N)_{\rm L} \times U(M)_{\rm R} 
\to U(N-M)_{\rm L} \times U(M)_{\rm V}$. 
Here, $U(N-M)_{\rm L}$ is the group generated by 
$\left( \pmatrix {{\bf 0}_M & {\bf 0}\cr
                  {\bf 0}   & T},
        {\bf 0}_M  \right)
\in ({\cal G}_{\rm L},{\cal G}_{\rm R})$, 
where $T$ are $(N-M) \times (N-M)$ matrices,  
and $U(M)_{\rm V}$ is generated by 
$\left( \pmatrix {     T  & {\bf 0}\cr
                  {\bf 0} & {\bf 0}_{N-M}},
                T  \right)
\in ({\cal G}_{\rm L},{\cal G}_{\rm R})$, 
where $T$ are $M \times M$ matrices.
The complex isotropy $\hat{\cal H}\pri$ that leaves 
$\left< \Ph \right>$ invariant is 
larger than ${\cal H\pri}^{\bf C}$ by 
$E \defeq \left( \pmatrix {{\bf 0}_M & T\cr
                  {\bf 0}   & {\bf 0}_{N-M}},
                 {\bf 0}_M  \right)$, 
where $T$ are $M \times (N-M)$ matrices.
Here, these $E$ constitute a Borel subalgebra ${\cal B}$ 
of $\hat{\cal H}\pri$, 
and its dimension is $\dim_{\bf C} {\cal B} = M(N-M)$. 
On the other hand, 
the complex broken generators consist of 
non-hermitian generators, 
$\bar E \defeq 
\left( \pmatrix { {\bf 0}_M & 0\cr
                        T   & {\bf 0}_{N-M} },
       {\bf 0}_M  \right)$, 
which are hermitian conjugates of $E$,  
and hermitian generators, 
$X \defeq 
\left( \pmatrix {  T      & {\bf 0}\cr
                  {\bf 0} & {\bf 0}_{N-M} },
                 - T  \right)$,  
which are elements of an axial symmetry $U(M)_{\rm A}$. 
The target manifold is a complex coset manifold
$M\pri \simeq {G\pri}^{\bf C}/\hat H\pri$, 
and its representative is 
$\xi\pri = \exp (\ph \cdot \bar{E} + i\psi \cdot X)
\defeq ({\xi\pri}_{\rm L},{\xi\pri}_{\rm R})$. 
The field can be written as 
$\vec{\phi} = \xi\pri \cdot V 
= {\xi\pri}_{\rm L} V {{\xi\pri}_{\rm R}}^{-1}$.
Its form near the vacuum is  
\beq
 \delta V 
 = \pmatrix{2i\psi\cr
            \ph }. \label{Gra-MP}
\eeq
Here, $\psi$ is an $M \times M$ matrix chiral superfield 
considered as mixed types and 
$\ph$ is an $(N-M)\times M$ matrix chiral superfield 
considered as pure types.
Hence, the numbers of mixed-type and 
pure-type chiral superfields are
$N_{\rm M} = M^2$ and 
$N_{\rm P} = M (N-M) (= \dim_{\bf C} {\cal B} )$, 
respectively.

\bigskip
To absorb the $M^2$ mixed-type chiral superfields, 
we gauge $U(M)_{\rm R}$ by introducing 
$M^2$ vector superfields
$V = V^A T_A$, where $T_A$ represents 
generators of $U(M)_{\rm R}$.
The gauged K\"{a}hler potential is
\beq
 K (\Ph,\Ph\dagg,V) = \tr(\Ph\dagg\Ph e^V) - c\, \tr V, 
     \label{gauged-Kahler2}
\eeq
where $c\, \tr V$ is a Fayet-Iliopoulos D-term. 
Since the vector superfields are transformed as
\beq
 e^V \to e^{V\pri} 
   = g_{\rm R} e^V {g_{\rm R}}\dagg,
\eeq
the gauged K\"{a}hler potential is invariant 
under the complexified gauge symmetry ${G_{\rm R}}^{\bf C}$.
To eliminate vector superfields, 
we use the equation of motion of $V$,\footnote{
We treat $e^{-V}\delta e^V$ as an infinitesimal 
parameter, since $\delta \tr(\Ph\dagg\Ph e^V) 
= \tr(\Ph\dagg\Ph e^V (e^{-V}\delta e^V))$. 
The second term is obtained from 
$\tr (\delta \log X )= \tr (X^{-1} \delta X)$, 
where $X = e^V$.}
\beq
 \delta K / \delta V = \Ph\dagg\Ph e^V - c {\bf 1}_M = 0 .
\eeq
Then $V$ can be solved as 
\beq
 V(\Ph,\Ph\dagg) = - \log \left({\Ph\dagg\Ph \over c}\right).
\eeq
By substituting this into Eq.~(\ref{gauged-Kahler2}), 
we obtain 
\beq
 K (\Ph,\Ph\dagg,V(\Ph,\Ph\dagg))= c\,\tr \log (\Ph\dagg\Ph)
 = c \log \det (\Ph\dagg\Ph), \label{Gra-Kahler}
\eeq
where we have omitted constant terms.
We choose the gauge fixing as  
\beq
 \Phi = \pmatrix {{\bf 1}_M \cr 
                  \ph},  \label{Gra-phi}
\eeq
where $\ph$ is an $(N-M)\times M$ matrix-valued chiral superfield.
By comparing Eq.~(\ref{Gra-MP}) and Eq.~(\ref{Gra-phi}), 
we find that all mixed-type multiplets $\psi$ 
have disappeared by this gauge fixing condition.
When $\xi$ is a representative of 
$\GC/\hat H=U(N)/U(N-M)\times U(M)$, 
$\Phi$ can be rewritten as
\beq
  \Phi = \xi \cdot V 
   = \xi_{\rm L} V {\xi_{\rm R}}^{-1},\hspace{0.5cm}
  \xi 
   = e^{\ph \cdot \bar E}
   = \left(
     \pmatrix {{\bf 1}_M & 0 \cr 
                \ph      &{\bf 1}_{N-M}} , 
     {\bf 1}_M \right)
   = (\xi_{\rm L}, \xi_{\rm R}). \label{Gra-phi2}
\eeq
Since the target space $M$ is parameterized solely by 
pure-type multiplets, 
it is a compact homogeneous K\"{a}hler manifold. 
By substituting Eq.~(\ref{Gra-phi}) into 
Eq.~(\ref{Gra-Kahler}), 
we obtain the K\"{a}hler potential of $M$:    
\beq
 K (\ph,\ph\dagg,V(\ph,\ph\dagg))
 = c \log \det ({\bf 1}_M + \ph\dagg\ph) .
\eeq
This is the K\"{a}hler potential of 
the Grassmann manifold 
$G_{N,M} = U(N)/U(N-M)\times U(M)$~\cite{Zu}.

\section{Nonlinear Sigma Models with F-term Constraints}
Only D-term constraints appeared in the last two examples.
In this section we also introduce appropriate F-term 
constraints to define other K\"{a}hlerian $G/H$ manifolds. 

\subsection{$SO(N)/SO(N-2) \times U(1)$}
We consider a global symmetry, 
$G\pri = SO(N)\times U(1)_{\rm D} = G\times U(1)_{\rm D}$. 
We will gauge $U(1)_{\rm D}$ symmetry later. 
The fields, which develop a vacuum expectation value, 
are $\vec{\phi}$ \ in the defining representation 
${\bf N}$ of $SO(N)$. 
The $U(1)_{\rm D}$ charge of $\vec{\phi}$ is defined to be $1$. 
The fundamental representation is decomposed 
under its subgroup $SO(N-2)\times U(1)$ as 
${\bf N} = ({\bf N-2},0) \oplus ({\bf 1},1)\oplus ({\bf 1},-1)$. 
Here, the second factor is the $U(1)$ charge.
The fields can be written as
\beq
 \vec{\phi} = \pmatrix{x \cr y^i \cr z}, 
\eeq
where $x$, $y^i\;(i=1,\cdots,N-2)$ and $z$ are 
a scalar, a vector and a scalar of $SO(N-2)$, respectively. 
Their $U(1)$ charges are defined above. 
$SO(N)$ is defined as the group that leaves 
the quadratic form
\beq
 I_2 \defeq \vec{\phi}^{\,2} 
  \defeq \vec{\phi}^{\,T} J \vec{\phi} = 2xz + y^2 
\eeq
invariant, where we have written the invariant tensor of rank 2 
in a rather unconventional way (see Appendix C): 
\beq
 J = \pmatrix{0 & {\bf 0}      & 1 \cr
         {\bf 0}& {\bf 1}_{N-2}&{\bf 0} \cr
              1 & {\bf 0}      & 0 } . \label{J-SO} 
\eeq
The generators of $SO(N)$ consist of 
the $SO(N-2)$ generator $T_{ij}\,(i,j=1,\cdots,N-2)$, 
the $U(1)$ generator $T$, and 
the upper triangular matrices $E^i \,(i=1,\cdots,N-2)$,  
which transform as $({\bf N-2},1)$, 
and their complex conjugates $\bar E_i = (E^i)\dagg$ 
in $({\bf N-2},-1)$. 
$SO(N)^{\bf C}$ acts on 
the fundamental representation in our basis as 
\beq
 \delta \vec{\phi}
 &=& \left(i \th T + {i \over 2} \th_{ij} T_{ij} 
    + \epb_i E^i + \ep^i \bar{E}_i \right) \vec{\phi} \non
 &=& \pmatrix 
 {   i\th &   \epb_j   & 0 \cr
    \ep^i &   \th_{ij} & - \epb_i \cr
        0 & - \ep^i    & -i\th}
 \pmatrix{x \cr y^j \cr z},  \label{SO-tr.law}
\eeq
where ${i \over 2}\th_{kl}{(T_{kl})^i}_j = \th_{ij}$. 
Here, these coefficients are normalized so that 
$\tr T_{ij}^2 = \tr T^2 = \tr E^i \bar{E}_i = 2$ (no sum). 
All parameters are complex when we consider $SO(N)^{\bf C}$ 
and real when we consider $SO(N)$. 

In order to impose the global symmetry 
$SO(N)\times U(1)_{\rm D}$, 
we introduce the superpotential 
\beq
 W(\phi_0,\vec{\phi}) = \phi_0 \vec{\phi}^{\,2},
\eeq
with the lagrange multiplier field $\phi_0$.
This is an $SO(N)$ singlet, and its $U(1)_{\rm D}$ charge 
is defined to be $-2$, so that $W$ is invariant under $G\pri$. 
Since the superpotential is a holomorphic function of 
$\phi$ and $\phi_0$, the symmetry is enhanced to 
its complexfication 
${G\pri}^{\bf C} = SO(N)^{\bf C}\times {U(1)_{\rm D}}^{\bf C}$. 
We can eliminate the auxiliary field 
by using its equation of motion,~\footnote{
There is another way to obtain the F-term constraint.
If we take 
$K = \lambda {\phi_0}\dagg\phi_0 
 + \vec{\phi}^{\,\dagger} \vec{\phi}$ and 
$W = \phi_0 \vec{\phi}^{\,2}$, then the potential reads 
$V = \1{\lambda} |\vec{\phi}^{\,2}|^2 
   + |\phi_0|^2 |\vec{\phi}|^2$. We obtain the F-term 
constraint in the limit $\lambda \to 0$.
}
\beq
 \del W /\del \phi_0 = I_2 
 = 2xz + y^2 = 0 .
\eeq
We thus obtain an F-term constraint ($N_F=1$). 
This equation is immediately solved to give  
\beq
 z = - {y^2 \over 2x}.
\eeq
Then, the field $\vec{\phi}$ constrained by the F-term 
can be written as 
\beq
 \vec{\phi} = \pmatrix{ x \cr y^i \cr -{y^2\over 2x} }.
\eeq
When this develops a vacuum expectation value, 
any vacuum can be transformed by ${G\pri}^{\bf C}$ to 
the standard form,  
\beq
 \vec{v} = \left<\vec{\phi}\right> 
  = \pmatrix{1 \cr {\bf 0}} .\label{SO-vev}
\eeq
By this vacuum expectation value, 
the global symmetry is spontaneously broken as 
$SO(N) \times U(1)_{\rm D}\to SO(N-2) \times U(1)\pri$, 
where the unbroken $U(1)\pri$ is generated by a linear 
combination of the $U(1)$ subgroup and $U(1)_{\rm D}$.\footnote{
Note that the condition $I_2=0$ is essential to introduce 
the gauge symmetry. To impose $I_2 = f^2 \neq 0$, 
we have to use $W = g \phi_0 (I_2 - f^2)$.
In this case there is no $U(1)_{\rm D}$ symmetry, 
and there is a supersymmetric vacuum alignment~\cite{Ni,Ni2}. 
Thus the unbroken global symmetry $H$ can 
depend on the choice of the vacuum 
expectation value: 
$H=SO(N-1)$ at the symmetric points, 
where $\phi\dagg\phi = f^2$, and 
$H=SO(N-2)$ at the non-symmetric points, 
where $\phi\dagg\phi > f^2$. 
Whereas $I_2=0$ corresponds to an open orbit, 
$I_2 \neq 0$ corresponds to closed orbits. 
In general, in closed orbits, 
there is a supersymmetric vacuum alignment. 
(See, e.g., Subsec.~3.3 for the $E_6$ case.) 
In this paper, we do not discuss closed orbits, 
since we cannot gauge the $U(1)_{\rm D}$ symmetry.
} 
The complex broken generators consist of $X$, 
which is hermitian and generates a mixed-type multiplet, 
and the $E^i$, which are non-hermitian and 
generate pure-type multiplets. 
Then, the number of the mixed- and pure-type multiplets are 
$N_{\rm M} =1$ and $N_{\rm P}= N-2$, respectively.
The target manifold $M\pri$ generated by these generators 
is non-compact and non-homogeneous due to 
the presence of the QNG boson.
The field near $\vec{v}$ is 
\beq
 \delta \vec{v} 
  = (i \psi X + \ph^i \bar{E}_i) \vec{v} 
  = \pmatrix{i \psi \cr \ph^i \cr 0} ,
\eeq
where $\psi$ is a mixed-type multiplet 
and $\ph^i$ are pure-type multiplets.

\bigskip
We elevate $U(1)_{\rm D}$ to a local gauge symmetry 
to obtain a compact manifold by eliminating the 
mixed-type multiplet, as in the case of ${\bf C}P^{N-1}$.
The gauged K\"{a}hler potential is 
the same as Eq.~(\ref{gauged-Kahler}).   
By integrating out the auxiliary superfields, 
we obtain Eq.~(\ref{Kahler}), 
with the constraint $\vec{\phi}^{\,2}=0$.  
By using the gauge symmetry ${U(1)_{\rm D}}^{\bf C}$, 
we can choose the gauge fixing as $x=1$: 
\beq
 \vec{\phi} = \pmatrix{ 1 \cr \ph^i \cr -\1{2}\ph^2 } . 
\eeq
Here we have rewritten $y^i$ as $\ph^i$.
This $\vec{\phi}$ can be rewritten 
by using the representative $\xi$ of 
the complex coset manifold 
$G^{\bf C}/\hat H = SO(N)/SO(N-2)\times U(1)$ as  
\beq
 \vec{\phi} = \xi \vec{v},\hspace{1cm}
 \xi = e^{\ph \cdot \bar{E}} 
     = \pmatrix{
             1 & {\bf 0}      & 0 \cr
         \ph^i & {\bf 1}_{N-2}& {\bf 0}\cr
   -\1{2}\ph^2 &  -\ph^i      & 1 } .\label{SO-phi}
\eeq
We thus obtain a K\"{a}hler potential of $G^{\bf C}/\hat H$, 
\beq
 K(\ph,\ph\dagg,V(\ph,\ph\dagg)) 
   = c \log \left(1 + |\ph|^2 
                + \1{4}\ph^{\dagger2} \ph^2 \right) \;.
\eeq
This is exactly the K\"{a}hler potential of 
$SO(N)/SO(N-2)\times U(1)$~\cite{DV,El,Ni2}. 

In our derivation of the K\"{a}hler potential, we used 
the D-term constraint after imposing the F-term constraint first.
Instead, we could impose the D-term constraint first. 
If we do so, we obtain the previous ${\bf C}P^{N-1}$ model. 
The F-term constraint is used as the holomorphic embedding 
condition of $Q^{N-2}({\bf C})= SO(N)/SO(N-2)\times U(1)$ 
to ${\bf C}P^{N-1}$.
It is a well-known method to obtain $Q^{N-2}({\bf C})$ 
in the mathematical literature~\cite{CV}. 
(See also p.~278 of Ref.~\cite{KN}.)

\subsection{$SO(2N)/U(N)$ and $Sp(N)/U(N)$}
In this subsection, 
we consider the global symmetry 
$G\pri = G_{\rm L}\times G_{\rm R}$, 
where $G_{\rm L}$ is either $SO(2N)$ or $Sp(N)$ 
and $G_{\rm R} = U(N)_{\rm R}$, 
which will be gauged later.
To embed $G_{\rm L}$ into a $2N\times 2N$ matrix of $U(2N)$, 
we write its elements by using four $N\times N$ matrices:
\beq
 g = \pmatrix {A & B\cr 
               C & D} \in U(2N) .
\eeq
This is an element of $SO(2N)$ or $Sp(N)$ if it satisfies
\beq
 g^T J\pri g = J\pri, \label{J-inv.}
\eeq
where $J\pri$ is the invariant tensor of $SO(2N)$ or $Sp(N)$: 
\beq
 J\pri = \pmatrix {{\bf 0}&{\bf 1}_N \cr
       \epsilon{\bf 1}_N &{\bf 0} }. \label{J}
\eeq
Here $\epsilon = +1$ corresponds to $SO(2N)$ and 
$\epsilon = -1$ to $Sp(N)$. 
Equation (\ref{J-inv.}) can be written explicitly as
\beq
 \pmatrix {A^T C + \epsilon C^T A & A^T D + \epsilon C^T B\cr 
           B^T C + \epsilon D^T A & B^T D + \epsilon D^T C }
 = \pmatrix {{\bf 0}&{\bf 1}_N \cr
       \epsilon{\bf 1}_N &{\bf 0} } .\label{J-inv.2} 
\eeq 

\bigskip
We consider the global symmetry as either
$G\pri =SO(2N)_{\rm L} \times U(N)_{\rm R}$ for $\epsilon = +1$ 
or
$G\pri =Sp(N)_{\rm L} \times U(N)_{\rm R}$ for $\epsilon = -1$.
The field content is
$\Phi \in ({\bf 2N},{\bf \bar N})$, 
which acquires a vacuum expectation value. 
Its transformation law under $G\pri$ is 
\beq
 \Phi \to \Phi\pri = g\cdot \Phi 
 = g_{\rm L} \Phi {g_{\rm R}}^{-1},\hspace{1cm} 
 g = (g_{\rm L},g_{\rm R}) \in G_{\rm L}\times G_{\rm R}
\eeq
The ${G\pri}^{\bf C}$ invariant superpotential is 
\beq
 W(\Phi_0,\Phi) = \tr (\Phi_0 \Phi^T J\pri \Phi) ,
  \label{SOSp-superpot.}
\eeq
where $\Phi_0$ is an $N\times N$ 
auxiliary matrix chiral superfield,  
whose transformation law is 
\beq
 \Phi_0 \to g_{\rm R} \Phi_0 {g_{\rm R}}^T .
\eeq
Since ${I_2}\pri \defeq \Phi^T J\pri \Phi$ 
is symmetric (anti-symmetric) 
for $\epsilon = 1$ ($\epsilon =-1$), $\Phi_0$ satisfies
\beq
 {\Phi_0}^T = \epsilon \Phi_0 .
\eeq
Hence, $\Phi_0$ belongs to a symmetric (anti-symmetric) 
rank-$2$ tensor representation 
of $SU(N)_{\rm R}$ for $\epsilon = 1$ ($\epsilon =-1$), 
and its $U(1)_{\rm D} (\in U(N)_{\rm R})$ charge is
defined to be $-2$ to cancel with the $\Phi$ charge.
Note that ${I_2}\pri = \Phi^T J\pri \Phi$ is 
invariant under $G_{\rm L}$, 
but not invariant under $G_{\rm R}$

To eliminate the auxiliary field $\Phi_0$, 
we solve its equation of motion 
\beq
 \delta W / \delta \Phi_0 = \Phi^T J\pri \Phi =0 .
  \label{SO,Sp-constraints}
\eeq
We thereby obtain F-term constraints for the fields $\Phi$. 
Their number is $N_F = \1{2}N (N+1)$ for $\epsilon = 1$ and 
$N_F = \1{2}N(N-1)$ for $\epsilon = -1$. 
Then the dimension of the resulting manifold $M\pri$ 
constrained by the F-term is 
$N_{\Phi} = 2N^2 - \1{2}N(N+1) = {3\over 2}N^2 -\1{2}N$ 
for $\epsilon = 1$ and 
$N_{\Phi} = 2N^2 - \1{2}N(N-1) = {3\over 2}N^2 + \1{2}N$ 
for $\epsilon = -1$.
When the field $\Phi$ acquires a vacuum expectation value, 
any vacuum can be transformed by ${G\pri}^{\bf C}$ 
to the standard form,
\beq
 V = \left<\Phi\right> 
   = \pmatrix{{\bf 1}_N \cr
              {\bf 0}_N} .
\eeq
Hence, the F-term constrained manifold 
is a ${G\pri}^{\bf C}$-orbit of $V$.
The breaking pattern of the global symmetry is either 
$SO(2N)_{\rm L} \times U(N)_{\rm R} \to U(N)_{\rm V}$ 
for $\epsilon =1$ or 
$Sp(N)_{\rm L} \times U(N)_{\rm R} \to U(N)_{\rm V}$ 
for $\epsilon =-1$.
Here, in both cases, the element of $U(N)_{\rm V}$ 
can be written as  
\beq
 \left( \pmatrix{ h & 0 \cr
                  0 & {h^{-1}}^T}, h \right) \in U(N)_{\rm V},
\eeq
where we have used Eq.~(\ref{J-inv.2}).
The complex isotropy group $\hat H\pri$ 
consists of complex extension of 
these elements and elements of the type
\beq
 \left(\pmatrix{{\bf 1}_N & B \cr 
                {\bf 0}_N & {\bf 1}_N},
        {\bf 1}_N \right) \defeq e^E,
  \hspace{1cm}
 E = \left(\pmatrix{{\bf 0}_N & B \cr 
                    {\bf 0}_N & {\bf 0}_N},
        {\bf 0}_N \right),   
\eeq
with the constraints, from Eq.~(\ref{J-inv.2}),
\beq
 B + \epsilon B^T = 0 .
\eeq
These $E$ constitute a Borel subalgebra ${\cal B}$ 
of $\hat{\cal H}\pri$.   
The dimensionality of ${\cal B}$ is 
$\dim_{\bf C} {\cal B} = \1{2}N(N-1)$ for $\epsilon = +1$ 
and $\dim_{\bf C} {\cal B} = \1{2}N(N+1)$ for $\epsilon = -1$. 
The pure-type broken generators are 
the complex conjugation of $E \in {\cal B}$: 
$\bar E = (E)\dagg$. 

\bigskip
To obtain a compact coset manifold, 
we gauge the $U(N)_{\rm R}$ symmetry by 
introducing vector superfields,  
as in the Grassmann manifold.
The gauged K\"{a}hler potential is the same as 
Eq.~(\ref{gauged-Kahler2}), but with F-term constraints.
Since the procedure of integrating out the gauge fields 
is also the same as for the Grassmann manifold, 
we obtain Eq.~(\ref{Gra-Kahler}). 
We can choose the gauge fixing as
\beq
 \Phi = \pmatrix {{\bf 1}_M \cr 
                  \ph},  \label{SO,Sp-phi}
\eeq
where $\ph$ satisfies the F-term constraints 
Eq.~(\ref{SO,Sp-constraints}): 
\beq 
 \Phi^T J\pri \Phi =\ph + \epsilon \ph^T = 0. 
  \label{SO,Sp-const2}
\eeq
The fields $\ph$ are all pure-type chiral superfields, 
since $\Phi$ is generated by 
the pure-type broken generators $\bar E$ from the vacuum $V$: 
\beq
 \Phi = \xi \cdot V,\hspace{1cm} \xi 
 = e^{\ph \cdot \bar{E}} = 
 \left(\pmatrix{{\bf 1}_N & {\bf 0}_N\cr 
                      \ph & {\bf 1}_N},
        {\bf 1}_N \right) .\label{SO,Sp-phi2}
\eeq
Here, from Eq.~(\ref{J-inv.2}), 
$\ph$ satisfies $\ph + \epsilon \ph^T =0$, 
which is consistent with (\ref{SO,Sp-const2}). 
By substituting Eq.~(\ref{SO,Sp-phi}) into  
Eq.~(\ref{Gra-Kahler}), 
we obtain the K\"{a}hler potential
\beq
 K (\ph,\ph\dagg,V(\ph,\ph\dagg))
 = c \log \det ({\bf 1}_N + \ph\dagg\ph),\hspace{1cm}
 \ph + \epsilon \ph^T = 0 .
\eeq
The fields $\ph$ are anti-symmetric (symmetric) parts of 
the matrix chiral superfield 
of the Grassmann manifold 
$G_{2N,N}$ for $\epsilon = +1$ ($-1$). 
Their dimensions are 
$\dim_{\bf C} M = \1{2} N(N-1)$ for $\epsilon = +1$ and 
$\dim_{\bf C} M = \1{2} N(N+1)$ for $\epsilon = -1$. 
Again, it is well-known that 
these manifolds are submanifolds of the Grassmann manifold 
$G_{2N,N}$ in the mathematical literature~\cite{CV}. 

\subsection{$E_6/SO(10) \times U(1)$}
This and the next subsections are devoted to 
the gauge theory construction of 
exceptional-type hermitian symmetric spaces.
The situation here is slightly different from 
the classical group cases. 
Namely, in the present case, 
an F-term constrained manifold $M\pri$ 
is characterized by the derivative of a $G$-invariant 
($\del I = 0$), but not the $G$-invariant itself ($I=0$), 
as in the case of classical types.

As in the $Q^{N-2}({\bf C})$ case, 
we consider the global symmetry 
$G\pri = E_6 \times U(1)_{\rm D} = G\times U(1)_{\rm D}$.
The field belongs to the fundamental 
representation of $E_6$: 
$\vec{\phi} \in {\bf 27}$, 
which will acquire a vacuum expectation value. 
We decompose $E_6$ under 
its maximal subgroup $SO(10) \times U(1)$.  
Since the fundamental representation can be decomposed as 
${\bf 27} = ({\bf 1},4) \oplus ({\bf 16},1) 
\oplus ({\bf 10},-2)$~\cite{Sl}, where the second entries 
are the $U(1)$ charges, 
the basic field $\vec{\phi}$ can be written as
\beq 
 \vec{\phi} = \pmatrix{ x \cr y_{\alpha} \cr z^A}.
\eeq
Here, $x$, $y_{\alpha}$ ($\alpha=1,\cdots,16$) and 
$z^A$ ($A=1,\cdots,10$) are an $SO(10)$ scalar, 
a Weyl spinor and a vector, respectively. 
The decomposition of the adjoint representation, 
${\bf 78} = ({\bf 45},0) \oplus ({\bf 1},0)
\oplus({\bf 16},1)\oplus ({\bf \bar{16}},-1)$~\cite{Sl}, 
implies that the $E_6$ algebra can be constructed 
with the $SO(10)$ generators $T_{AB}$ $(A,B=1,\cdots,10)$,  
the $U(1)$ generator $T$, upper half generators $E_{\alpha}$, 
which belong to a Weyl spinor of $SO(10)$,  
and their conjugates $\bar E^{\alpha}$. 
(See Appendix D for details.) 

The transformation law of $\vec{\phi}$ under 
the complex extension of $E_6$ is~\cite{IKK2,KuSa}
\beq
 \delta \vec{\phi}
 &=& \left(i \th T + {i \over 2} \th_{AB} T_{AB} 
    + \epb^{\alpha} E_{\alpha} 
    + \ep_{\alpha} \bar{E}^{\alpha} \right) 
  \vec{\phi} \non
 &=& \pmatrix 
 {{2i \over \sqrt{3}} \th & \epb^{\beta} & {\bf 0}\cr
  \ep_{\alpha} & 
    {i\over 2} \th_{AB}{(\sigma_{AB})_{\alpha}}^{\beta} 
     +{i\over 2\sqrt{3}} \theta \delta_{\alpha}^{\beta} & 
        -\1{\sqrt 2} (\epb \sigma_B C)_{\alpha} \cr
   {\bf 0} & - \1{\sqrt 2} (C{\sigma_A}\dagg \ep)^{\beta} &
          \th_{AB} -{i \over \sqrt{3}}\th \delta_{AB} }
  \pmatrix{ x \cr y_{\beta} \cr z^B} , \label{E6-tr.}
\eeq
where ${i\over 2}\th_{CD} \rho{(T_{CD})^A}_B = \th_{AB}$, 
and $\rho(T_{AB})$ is the vector representation 
matrices of $SO(10)$. 
The $16 \times 16$ matrices $\sigma_A$, $\sigma_{AB}$ and $C$ are 
(off-diagonal blocks of) $SO(10)$ gamma matrices, 
spinor rotation matirices 
and the charge conjugation matrix, respectively. 
Normalizations are fixed by 
$\tr T^2 = \tr {T_{AB}}^2 
= \tr E_{\alpha} \bar{E}^{\alpha}=6$ (no sum).\footnote{
$\tr {T_{AB}}^2 = 6$ has been calculated from 
$\tr \rho(T_{AB})^2 =2$, while $\tr(\sigma_{AB})^2 =4$ 
and others have been fixed to this.
}

\bigskip 
The decomposition of the tensor product, 
${\bf 27} \otimes {\bf 27} = {\bf \bar{27}}_{\rm s} \oplus \cdots$, 
implies that there exist a rank-$3$ symmetric 
invariant tensor $\Gamma_{ijk}$ and 
its complex conjugate $\Gamma^{ijk}$~\cite{KuSa}. 
By using this invariant tensor, 
a cubic invariant $I_3$ of $E_6$ is defined as
\beq
 I_3 \defeq \Gamma_{ijk} \phi^i \phi^j \phi^k 
  = x z^2 + \1{\sqrt 2} z^A (y C {\sigma_A}\dagg y) .
\eeq
Note that this is not invariant under $U(1)_{\rm D}$.

We construct the superpotential 
\beq
 W (\vec{\phi}_0, \vec{\phi}) 
 = \Gamma_{ijk} {\phi_0}^i \phi^j \phi^k .
\eeq
Here $\vec{\phi}_0$ represents auxiliary fields whose 
$U(1)_{\rm D}$ charges should be chosen so as to make 
the superpotential invariant. 
If we assign the 
$U(1)_{\rm D}$ charge $1$ to $\vec{\phi}$, $\vec{\phi}_0$
must have charge $-2$, so that they belong to $({\bf 27},-2)$. 
The equations of motion for the auxiliary fields ${\phi_0}^i$, 
$\delta W /\delta \phi_0 = \Gamma_{ijk} \phi^j \phi^k = 0$, 
are 
\beq
 &&\del W/\del {z_0}^A 
 = \Gamma_{Ajk} \phi^j \phi^k 
 =  2 z_A x + \1{\sqrt 2}  
    y_{\alpha} (C {\sigma_A}\dagg)^{\alpha\beta} y_{\beta}
 = 0 , \\ 
 &&\del W/\del y_{0\alpha} 
   = \Gamma_{\alpha jk} \phi^j \phi^k 
   = \sqrt 2 (C {\sigma_A}\dagg)^{\alpha\beta} y_{\beta} z^A  
   = 0 , \\
 &&\del W/\del x_0 
   = \Gamma_{0jk} \phi^j \phi^k 
   = z^2 = 0 .
\eeq
In the second equation, we have used the fact that 
$(C {\sigma_A}\dagg)^{\alpha\beta}$ is symmetric.
Note that these equations can also be written as 
\beq 
 \del I_3=0,
\eeq 
where the differentiation is with respect to $\phi^i$. 
In these $27$ equations, 
only the first $10$ equations are independent.
The first equation can be solved to yield
\beq
 z_A = -\1{2\sqrt{2} x} y( C {\sigma_A}\dagg) y . 
         \label{E6-constraint}
\eeq
Then, the last two equations are not independent, 
since they are automatically satisfied as
\beq
 &&  \sqrt 2 (C {\sigma_A}\dagg)^{\alpha\beta} y_{\beta} z^A  
   = - \1{2x}(C {\sigma_A}\dagg)^{\alpha\beta} y_{\beta}
       \left(y (C {\sigma_A}\dagg)y \right) 
   = 0 , \\
 &&
 z^2 = \1{8x^2} \left(y (C {\sigma_A}\dagg) y \right)^2
  = 0 ,
\eeq
with the help of the identity
\beq
 (\varepsilon C {\sigma_A}\dagg \psi)(\psi C {\sigma_A}\dagg \eta)
 = -\1{2} (\varepsilon C {\sigma_A}\dagg \eta)
         (\psi C {\sigma_A}\dagg \psi)  .\label{spinor-id.}
\eeq
Hence, the number of F-term conditions is $N_F = 10$, 
and the dimension of $M\pri$ is $N_{\Phi}=27-10=17$. 
The manifold $M\pri$ satisfying these F-term constraints 
can be written as
\beq
 \vec{\phi}  
 = \pmatrix { x \cr
             y_{\alpha} \cr
              -\1{2 \sqrt 2x}(y C {\sigma_A}\dagg y)} .
\eeq
On $M\pri$, the value of the $E_6$ invariant is 
\beq
 I_3 \sim (y C {\sigma_A}\dagg y)^2 = 0,  
\eeq
by the identity~(\ref{spinor-id.}).
Note that $I_3$ must vanish, since it is not invariant 
under $U(1)_{\rm D}$.

When the fields $\vec{\phi}$ develop 
a vacuum expectation value, 
any vacuum can be transformed under ${G\pri}^{\bf C}$ to 
the standard form, 
\beq
 \vec{v} = \left<\vec{\phi}\right> 
         = \pmatrix{1 \cr {\bf 0}}. \label{vev-E6}
\eeq
The global symmetry is spontaneously broken as
$E_6 \times U(1)_{\rm D} \to SO(10) 
\times U(1)\pri = H\pri$.~\footnote{
As in the case of $SO(N)$ discussed in Subsec.~3.1, 
there is no $U(1)_{\rm D}$ symmetry if $I_3 \neq 0$. 
In this case, the ${E_6}^{\bf C}$-orbit is closed, and, 
by a supersymmetric vacuum alignment, 
there exist two regions with different 
unbroken global symmetries~\cite{Ni}, 
symmetric points and non-symmetric points.  
The breaking patterns of $E_6$ are 
$E_6 \to F_4$ at the symmetric points and $E_6 \to SO(8)$ 
at generic points~\cite{KuSa}.
} 
The unbroken $U(1)\pri$ is generated by 
$T\pri = T - {2\over \sqrt{3}}{\bf 1}_{27} 
= {\rm diag}.(0,-{\sqrt{3} \over 2} \delta_{\alpha}^{\beta}, 
              -\sqrt{3}\delta_{AB})$, 
and $SO(10)$ is generated by $T_{AB}$. 
The complex isotropy $\hat {\cal H}\pri$ 
is larger than the complexification of ${\cal H}\pri$
due to the existence of the $E_{\alpha}$. 
These 16 $E_{\alpha}$ constitute a Borel 
subalgebra ${\cal B}$ in $\hat {\cal H}\pri$.
The complex broken generators are composed of 
pure-type generators $\bar E^{\alpha}$ and 
another combination of $U(1)$ generators of a mixed-type 
$X \sim (1,\cdots)$. 
Their numbers are $N_{\rm P} = 16$ and 
$N_{\rm M} = 1$, respectively.
The target manifold $M\pri$ generated 
by these broken generators has dimension 
$\dim M\pri = N_{\Phi}= 17$. 
Since this coincides with the dimension of the manifold 
constrained by the 10 independent F-term conditions, 
any vacuum that satisfies F-term constraints 
can be transformed to the form of Eq.~(\ref{vev-E6}) 
by a ${G\pri}^{\bf C}$ transformation.

\bigskip
To remove the mixed-type multiplet and 
to obtain a compact manifold,
we gauge the $U(1)_{\rm D}$ symmetry 
as in the case of ${\bf C}P^{N-1}$.
The gauged K\"{a}hler potential is the same as 
in Eq.~(\ref{gauged-Kahler}).
Since the procedure to eliminate the vector superfield 
is also the same as in the ${\bf C}P^{N-1}$ case, 
we obtain Eq.~(\ref{Kahler}). 
We can choose a gauge fixing as 
\beq
 \vec{\phi}  
 = \pmatrix { 1 \cr
              \ph_{\alpha} \cr
              -\1{2 \sqrt 2}(\ph C {\sigma_A}\dagg \ph)},
   \label{E6-gauge}
\eeq
where we write $\ph_{\alpha}$ for $y_{\alpha}$.
By using the representative $\xi$ 
of the complex coset manifold 
$M=\GC/\hat H \simeq E_6/SO(10)\times U(1)$, 
$\vec{\phi}$ can be rewritten as
\beq
 \vec{\phi} = \xi \vec{v}, \hspace{1cm} 
 \xi = e^{\ph \cdot \bar{E}}
 =  \pmatrix 
 { 1            & {\bf 0}       & {\bf 0} \cr
   \ph_{\alpha} & {\bf 1}_{16}  & {\bf 0} \cr
   -{1 \over 2\sqrt 2} (\ph C{\sigma_A}\dagg \ph )
      & -{1 \over \sqrt 2} 
         (C{\sigma_A}\dagg \ph)^{\beta} & {\bf 1}_{10}}. 
    \label{E6-phi}  
\eeq
By substituting Eq.~(\ref{E6-gauge}) into 
Eq.~(\ref{Kahler}), we obtain the K\"{a}hler potential 
\beq
 K(\ph,\ph\dagg,V(\ph,\ph\dagg)) 
 = c \log \left( 1 + |\ph|^2 + 
 \1{8} (\ph\dagg {\sigma^A}\ph\dagg)
       (\ph {\sigma_A}^{\dagger} \ph)\right) ,
\eeq
where we have used the basis in which $C=1$~\cite{IKK2}.
This coincides with the K\"{a}hler potential of 
$E_6/SO(10)\times U(1)$ constructed 
in Refs.~\cite{CV,IKK2,El}. 
(It is also equivalent to Ref.~\cite{AAH}.)
Its dimension is $\dim_{\bf C} M = 27 - 10 -1 = 16$.
If we do not introduce the superpotential, 
the manifold is ${\bf C}P^{26}$. 
Hence, $E_6/SO(10)\times U(1)$ 
is embedded in ${\bf C}P^{26}$ 
by $10$ F-term constraints, $\del I_3 = 0$.
In fact, Yasui constructed $E_6/SO(10)\times U(1)$ as 
a submanifold of ${\bf C}P^{26}$ by 
using the Jordan algebra~\cite{Ya}.

\subsection{$E_7/E_6 \times U(1)$}
In this subsection, 
we consider another exceptional group, $E_7$. 
The global symmetry in this case is 
$G\pri = E_7 \times U(1)_{\rm D} = G\times U(1)_{\rm D}$.
The basic fields $\vec{\phi}$ belong to the fundamental 
representation ${\bf 56}$. 
Under a maximal subgroup $E_6 \times U(1)$, 
this representation can be decomposed as
${\bf 56} = ({\bf 27}, -\1{3})\oplus ({\bf \bar{27}},\1{3}) 
\oplus ({\bf 1},-1) \oplus ({\bf 1},1)$~\cite{Sl}. 
Therefore, we write $\vec{\phi}$ as 
\beq
 \vec{\phi} = \pmatrix{ x \cr y^i \cr z_i\cr w},  
\eeq
where $y^i$ and $z_i$ are ${\bf 27}$ and ${\bf \bar{27}}$, 
respectively, 
and $x$ and $w$ are scalars.
By a decomposition of the adjoint representation 
under $E_6 \times U(1)$~\cite{Sl}, 
${\bf 133} = ({\bf 78},0) \oplus ({\bf 1},0)
\oplus ({\bf 27},1) \oplus ({\bf \bar{27}},-1)$, 
we can construct the $E_7$ algebra from 
the $E_6$ algebra $T_A\,(A=1,\cdots,78)$, 
the $U(1)$ generator $T$, 
the upper half generators $E^i \,(i=1,\cdots,27)$, 
belonging to ${\bf 27}$, 
and their conjugates $\bar E_i = (E^i)\dagg$, 
belonging to ${\bf \bar{27}}$. 
(Their commutation relations are discussed 
in Appendix E.)  
The action of the $E_7$ algebra on 
the fundamental representation is 
\beq
 \delta \vec{\phi}
 &=& \left(i \th T + i \th_A T_A 
    + \epb_i E^i + \ep^i \bar{E}_i \right) 
  \vec{\phi} \non
 &=& \pmatrix 
 { i\sqrt{3\over 2}\th & \epb_j  & {\bf 0} & 0\cr
     \ep^i & i\th_A {\rho(T_A)^i}_j 
            + i\sqrt{\1{6}}\theta {\delta^i}_j & 
                           {\Gamma}^{ijk} \epb_k & {\bf 0}\cr
  {\bf 0} & \Gamma_{ijk}\ep^k & - i\th_A {{\rho({T_A})^T}_i}^j 
          - i\sqrt{\1{6}}\theta {\delta_i}^j & \epb_i\cr
   0 & {\bf 0} & \ep^j & -i\sqrt{3\over 2}\th }
 \pmatrix{ x \cr y^j \cr z_j\cr w} ,\non \label{E7_action}
\eeq
where $\rho(T_A)$ is the $27\times 27$ representation matrix 
for the fundamental representation, 
$\Gamma_{ijk}$ is the $E_6$ invariant tensor, 
defined in the last subsection, 
and $\Gamma^{ijk}$ is its conjugate.
Here normalizations have been determined by 
$\tr T^2 = \tr {T_A}^2 = \tr E^i \bar{E}_i = 12$ 
(no sum).\footnote{
$\tr {T_A}^2 = 12$ has been calculated with 
the normalization $\tr (\rho(T_A)^2 )= 6$
for the $E_6$ fundamental representation,  
as in the previous subsection.
Other normalizations have been fixed relative to this. 
In the calculation of $\tr E^i \bar{E}_i =12$, 
we have used the identity Eq.~(\ref{E6-id.}).
}

In the tensor products~\cite{Sl} 
${\bf 56} \otimes {\bf 56} = {\bf 1}_{\rm a} \oplus \cdots$ and  
${\bf 56} \otimes {\bf 56}\otimes {\bf 56}\otimes {\bf 56} 
= {\bf 1}_{\rm s} \oplus \cdots$, 
there exist the rank-$2$ anti-symmetric invariant tensor 
$f_{\alpha\beta}$ and 
the rank-$4$ symmetric invariant tensor 
$d_{\alpha\beta\gamma\delta}$, respectively. 
Their components are calculated in Appendix E. 
By using this invariant tensor, 
we can construct the quartic invariant of $E_7$ as 
\beq
 I_4 &\defeq& d_{\alpha\beta\gamma\delta} 
     \phi^{\alpha} \phi^{\beta} \phi^{\gamma} \phi^{\delta}\non
  &=& -\1{2} (x w - y^i z_i)^2 
   -\1{3}w \Gamma_{ijk} y^iy^jy^k -\1{3}x \Gamma^{ijk} z_iz_jz_k\non
   &&+\1{2} \Gamma^{ijk}\Gamma_{ilm} z_j z_k y^l y^m .  
\eeq
Again, note that this is not invariant under $U(1)_{\rm D}$.

The superpotential invariant under 
$E_7 \times U(1)_{\rm D}$ is 
\beq
 W (\vec{\phi}_0, \vec{\phi}) 
  = d_{\alpha\beta\gamma\delta} 
  {\phi_0}^{\alpha} \phi^{\beta} \phi^{\gamma} \phi^{\delta} ,
\eeq
where the ${\phi_0}^{\alpha}$ are auxiliary fields 
belonging to $({\bf 56},-3)$. 
Here the second component is the $U(1)_{\rm D}$ charge 
assigned to cancel the  $U(1)_{\rm D}$ charge of $\phi^{\alpha}$.
(The term with rank-$2$ tensor $f_{\alpha\beta}$ 
is forbidden by $U(1)_{\rm D}$ symmetry.) 
To eliminate the auxiliary fields $\phi_0$, 
we consider F-term constraints 
obtained from their equation of motions: 
\beq
 &&\del W /\del {y_0}^i
 = w (x z_i - \Gamma_{ijk} y^j y^k) - z_i y^j z_j
  + \Gamma^{jkl} \Gamma_{jim} z_k z_l y^m = 0 ,\non
 &&\del W /\del w_0
 = x y^i z_i - w x^2 -\1{3}\Gamma_{ijk}y^i y^j y^k = 0 ,\non
 &&\del W /\del z_{0 i} 
 =  x (w y^i - \Gamma^{ijk} z_j z_k) - y^i y^j z_j
  + \Gamma^{jik} \Gamma_{jlm} z_k y^l y^m = 0  ,\non
 &&\del W /\del x_0 
 = w y^i z_i - x w^2 -\1{3}\Gamma^{ijk}z_i z_j z_k = 0. 
 \label{E7_F-flat}
\eeq
Note that these equations can be written as 
\beq
 \del I_4 = 0,
\eeq
where the differentiations are with respect to $\phi^{\alpha}$. 
We show that only half of these $58$ 
equations are independent.
To solve these equations, we put the ansatz 
\beq
 z_i = {c \over x} \Gamma_{ijk} y^j y^k ,\label{ansatz}
\eeq
where $c$ is a constant. 
By substituting this ansatz into 
the first and second equations, we obtain 
\beq
 &&w (c-1) \Gamma_{ijk} y^j y^k 
 + {c^2 \over 3 x^2}\Gamma_{ijk}\Gamma_{lmn}y^jy^ky^ly^my^n =0 ,\\
 &&w = {c -\1{3} \over x^2} \Gamma_{ijk}y^i y^j y^k 
  \label{omega}.
\eeq
From these equations we obtain the equation 
\beq
 {4(c-\1{2})^2 \over 3x^2} 
 \Gamma_{ijk}y^jy^k \Gamma_{lmn}y^ly^my^n =0 , 
\eeq
which gives $c=\1{2}$. By substituting $c=\1{2}$ back 
into Eqs.~(\ref{ansatz}) and (\ref{omega}), 
we obtain the results, 
\beq
 z_i = \1{2 x} \Gamma_{ijk}y^j y^k ,\;\;\;
   w = \1{6 x^2} \Gamma_{ijk} y^i y^j y^k .\label{E7-ans.}
\eeq
In the same way, 
the third and the fourth equations 
in Eqs.~(\ref{E7_F-flat}) can be solved as
\beq
 y^i = \1{2 w} \Gamma^{ijk}z_j z_k ,\;\;\;
   x = \1{6 w^2} \Gamma^{ijk} z_i z_j z_k .
\eeq
We can show that 
these equations are not independent of Eqs.~(\ref{E7-ans.})
with the help of the Springer relation, Eq.~(\ref{Springer}).
Then the number of F-term constraints is $N_F=28$, 
and the dimension of $M\pri$ is 
$\dim_{\bf C}M\pri = 56 -28 =28$. 
Thus, the F-term constraints can be solved as
\beq
 \vec{\phi}
  = \pmatrix{ x \cr
              y^i \cr
             \1{2x} \Gamma_{ijk} y^j y^k  \cr
             \1{6x^2} \Gamma_{ijk} y^i y^j y^k} .
 \label{E7_F-flat2}
\eeq
On these points, the value of 
the $E_7$ invariant is 
\beq
 I_4 = 0, 
\eeq  
where we have used the Springer relation (\ref{Springer}).  
Note that $U(1)_{\rm D}$ invariance requires $I_4=0$.

By using ${G\pri}^{\bf C}$, 
any vacuum expectation value of $\vec{\phi}$ 
can be transformed to
\beq
 \vec{v} =\left<\vec{\phi}\right> = \pmatrix{1 \cr {\bf 0}}.
   \label{E7_VEV}
\eeq 
On this vacuum, global symmetry is spontaneously broken as 
$E_7 \times U(1)_{\rm D} \to E_6 \times U(1)\pri \defeq H\pri$.
Here $U(1)\pri$ is generated by a linear combination of 
the $U(1)$ generator $T$ and 
the $U(1)_{\rm D}$ generator $T_{\rm D} ={\bf 1}_{56}$. 
From Eq.~(\ref{E7_action}), 
we see that the complex isotropy $\hat {\cal H}\pri$ 
is larger than ${{\cal H}\pri}^{\bf C}$
due to the presence of the $E^i$, 
which constitute a Borel subalgebra.  
The complex broken generators constitute 
a hermitian generator $X$, 
which is a linear combination of $T_{\rm D}$ and $T$, 
and non-hermitian generators ${\bar E}_i$. 
Hence, the numbers of mixed- and pure-type multiplets are 
$N_{\rm M}=1$ and $N_{\rm P} = 27$, respectively.
The target manifold $M\pri$ is 
generated by these broken generators,  
and its dimension is $\dim_{\bf C} M\pri = 28$, 
which coincides with the dimension of the manifold 
constrained by the F-term conditions in Eq.~(\ref{E7_F-flat2}). 
 
\bigskip
The target manifold $M\pri$ obtained above is 
non-compact due to the QNG boson. 
We gauge the $U(1)_{\rm D}$ symmetry 
to remove the mixed-type multiplet and to obtain 
a compact manifold.
Since the situation is the same as for  
the ${\bf C}P^{N-1}$, $Q_{N-2}({\bf C})$ 
and $E_6/SO(10)\times U(1)$ cases, 
by integrating out the vector superfield, 
we obtain Eq.~(\ref{Kahler}).
We can choose the gauge fixing as
\beq
 \vec{\phi}
  = \pmatrix{ 1 \cr
             \ph^i \cr
             \1{2} \Gamma_{ijk} \ph^j \ph^k  \cr
             \1{6} \Gamma_{ijk} \ph^i \ph^j \ph^k  } ,
    \label{E7_gauge-fixing}
\eeq
where we rewrite $y^i$ as $\ph^i$.
As in the previous subsections, 
this can be written as
\beq
 \vec{\phi} = \xi \vec{v},\hspace{1cm} 
 \xi = e^{\ph \cdot \bar E} = \pmatrix 
 { 1     & {\bf 0}      & {\bf 0}     & 0\cr
  \ph^i  & {\bf 1}_{27} &{\bf 0}_{27} &{\bf 0}\cr
 \1{2}\Gamma_{ijk} \ph^j \ph^k  &
             \Gamma_{ijk}\ph^j&{\bf 1}_{27}&{\bf 0}\cr 
 \1{6}\Gamma_{ijk}\ph^i\ph^j\ph^k&
             \1{2}\Gamma_{ijk}\ph^j\ph^k& \ph^i& 1} .
  \label{E7-phi}
\eeq
Hence the target manifold $M$, 
obtained by integrating out the vector superfield, 
is the coset manifold generated by $\bar E_i$, 
which is $M \simeq E_7/E_6\times U(1)$. 
Then, by substituting (\ref{E7_gauge-fixing}) 
into Eq.~(\ref{Kahler}),
we obtain the  K\"{a}hler potential
\beq
 K(\ph,\ph\dagg,V(\ph,\ph\dagg)) 
 = c \log \left( 1 + |\ph|^2 + \1{4}|\Gamma_{ijk}\ph^j\ph^k|^2 
   + \1{36}|\Gamma_{ijk}\ph^i\ph^j\ph^k|^2\right) .
\eeq
This form coincides with Ref.~\cite{DV}.
Its dimension is $\dim_{\bf C} M = 56 - 28 -1 =27$.
It can be embedded into ${\bf C}P^{55}$ by 
holomorphic constraints $\del I_4 = 0$.

\section{Conclusions and Discussion} 
We have obtained nonlinear sigma models 
whose target manifolds are 
the hermitian symmetric spaces $G/H$, 
which are compact and homogeneous, from linear models. 
For this purpose, we introduced appropriate superpotentials 
for $G=SO, SU, Sp, E_6$ and $E_7$ to impose F-term constraints. 
By solving these F-term constraint equations, we have obtained 
constrained manifolds $M\pri$, 
which are non-compact and non-homogeneous 
due to the existence of QNG bosons. 
When there is no gauge symmetry, 
there must be at least one QNG boson,  
by the theorem of Lerche and Shore~\cite{LS}, and 
the manifold inevitably becomes non-compact and non-homogeneous 
(see Appendix B). 
In order to get rid of these unwanted QNG-bosons, we further 
introduced suitable local gauge symmetry. By choosing suitable 
gauge conditions, we obtained the K\"{a}hler potentials of 
all the hermitian symmetric spaces, where decay constants 
(overall constants of K\"{a}hler potentials) 
originate from FI-terms of gauge fields. 

The gauging procedures to eliminate QNG bosons 
can be summarized as follows:\footnote{
From the result in Ref.~\cite{Ni}, 
in all cases considered in this paper, 
we know that there exists no supersymmetric vacuum alignment, 
since there is no non-singlet broken generators under 
the real unbroken subgroup $H$. 
Hence, the F-term constrained manifolds 
$M\pri \simeq {G\pri}^{\bf C}/\hat H\pri$ 
are topologically isomorphic to 
direct products of a QNG boson factor 
${\bf R}^+ = \{ \th|\th \in {\bf R}, \th>0 \}$, 
which is non-compact,  
and a NG bosons factor $G\pri/H\pri$, 
which is compact. 
For example, 
in the case of ${\bf C}^{N}$ 
without an F-term constraint, 
$M\pri \simeq {G\pri}^{\bf C}/\hat H\pri 
= {(SU(N) \times U(1)_{\rm D})^{\bf C} \over 
(SU(N-1) \times U(1)\pri)^{\bf C} \wedge {\cal B}} 
\simeq {\bf R}^+ \times 
{SU(N) \times U(1)_{\rm D}\over 
SU(N-1) \times U(1)\pri}
= {\bf R}^+ \times {G\pri \over H\pri}$. 
Then, by gauging $U(1)$, 
we obtain $\GC/\hat H \simeq G/H = {\bf C}P^{N-1}$.
} 
\beq
  {\bf R}^+ \times 
 {SU(N) \times U(1)_{\rm D} \over SU(N-1) \times U(1)\pri} 
     &\stackrel{U(1)_{\rm D}}{\longrightarrow}& 
   {SU(N) \over S(U(N-1) \times U(1))}, \non
  ({\bf R}^+)^{M^2} \times
 {U(N)_{\rm L} \times U(M)_{\rm R} 
      \over U(N-M)_{\rm L} \times U(M)_{\rm V}} 
    &\stackrel{U(M)_{\rm R}}{\longrightarrow}&  
  {U(N)_{\rm L} \over U(N-M)_{\rm L} \times U(M)_{\rm L}}, \non
  {\bf R}^+ \times
 {SO(N) \times U(1)_{\rm D} \over SO(N-2) \times U(1)\pri}  
     &\stackrel{U(1)_{\rm D}}{\longrightarrow}& 
   {SO(N) \over SO(N-2) \times U(1)}, \non
  ({\bf R}^+)^{N^2}\times
 {SO(2N)_{\rm L} \times U(N)_{\rm R} \over U(N)_{\rm V}}
    &\stackrel{U(N)_{\rm R}}{\longrightarrow}& 
  {SO(2N)_{\rm L} \over U(N)_{\rm L}}, \non
  ({\bf R}^+)^{N^2} \times
 {Sp(N)_{\rm L} \times U(N)_{\rm R} \over U(N)_{\rm V}}
    &\stackrel{U(N)_{\rm R}}{\longrightarrow}& 
  {Sp(N)_{\rm L} \over U(N)_{\rm L}}, \non
  {\bf R}^+ \times
 {E_6 \times U(1)_{\rm D} \over SO(10)\times U(1)\pri}
    &\stackrel{U(1)_{\rm D}}{\longrightarrow}& 
  {E_6 \over SO(10) \times U(1)}, \non 
  {\bf R}^+ \times
 {E_7 \times U(1)_{\rm D} \over E_6\times U(1)\pri}
    &\stackrel{U(1)_{\rm D}}{\longrightarrow}& 
  {E_7 \over E_6 \times U(1)}.  \nonumber \nopagebreak[1]
\eeq
The left-hand sides denote 
the F-term constrained manifolds $M\pri$ \ 
(if there is a superpotential). 
All $M\pri$ are non-compact and non-homogeneous, 
due to the existence of QNG bosons 
represented by ${\bf R}^+$. 
This implies that they are scale factors. 
The arrows represent the gauging and 
the right hand sides denote the manifold $M$  
obtained by integrating out the vector superfields. 
The relation between $M$ and $M\pri$ is  
a K\"{a}hler quotient, $M=M\pri/G_{\rm gauge}^{\bf C}$. 
All $M$ are compact and homogeneous, 
since they are parameterized by only NG bosons.      
In the cases of ${\bf C}P^{N-1}$ and $G_{N,M}$,  
there are no F-term constraints. 
Other cases have $\GC$-invariants, 
superpotentials and F-term constraints,  
as summarized in Table~2. 
\begin{table}
\caption{\bf F-term constraints and embedding}
\begin{center}
\begin{tabular}{|c|c|c|c|c|}
 \noalign{\hrule height0.8pt}
  $G/H$ & $\GC$-invariants & superpotentials & constraints 
  & embedding \\
 \hline
 \noalign{\hrule height0.2pt}
  ${SO(N) \over SO(N-2)\times U(1)}$ 
      & $I_2 = \vec{\phi}^{\,T} J \vec{\phi}$     
      & $\phi_0 I_2$ & $I_2=0$ & ${\bf C}P^{N-1}$\\
  ${SO(2N)\over U(N)}$, ${Sp(N)\over U(N)}$ 
      & ${I_2}\pri =\Phi^T J\pri \Phi$ 
      & $\tr (\Phi_0 {I_2}\pri)$ 
      & ${I_2}\pri = 0$ & $G_{2N,N}$\\
  ${E_6\over SO(10)\times U(1)}$ 
      & $I_3=\Gamma_{ijk}\phi^i\phi^j\phi^k$ 
      & $\Gamma_{ijk}{\phi_0}^i\phi^j\phi^k$ 
      & $\del I_3 =0$ & ${\bf C}P^{26}$\\
  ${E_7\over E_6 \times U(1)}$ 
      & $I_4=d_{\alpha\beta\gamma\delta}
         \phi^{\alpha}\phi^{\beta}
         \phi^{\gamma}\phi^{\delta}$ 
      & $d_{\alpha\beta\gamma\delta}
         {\phi_0}^{\alpha}\phi^{\beta}
         \phi^{\gamma}\phi^{\delta}$  
      & $\del I_4 =0$ & ${\bf C}P^{55}$\\  
 \noalign{\hrule height0.8pt}
 \end{tabular}
 \end{center}
\begin{footnotesize}
Here, 
$J$, $J\pri$, $\Gamma$ and $d$ are 
rank-2, rank-2, rank-3 and rank-4 
invariant symmetric tensors of 
$SO(N)$, $SO(2N)$ or $Sp(N)$, $E_6$ and $E_7$, respectively, 
and $I_2$, ${I_2}\pri$, $I_3$ and $I_4$ are 
$\GC$-invariants composed of them. 
Each superpotential gives an F-term constraint, 
which is $I=0$ in the case of classical groups 
and $\del I=0$ in the case of exceptional groups. 
Only 10 equations of the 27 equations are 
independent in the $E_6$ case, 
and only 28 equations among 56 equations 
are independent in the $E_7$ case. 
The last column denotes the projective or Grassmann manifold, 
in which each hermitian symmetric space is embedded 
by the F-term constraint. 
\end{footnotesize}
\end{table}

The F-term constraints can be classified 
into two types: 
\begin{itemize}
\item
$G=SO,\; Sp$: $I=0$. (They are $\GC$-invariant.) 
\item
$G= E_6,\; E_7$: $\del I = 0$. 
(Although the $\del I$ are not $\GC$-invariant, 
the constraints themselves are $\GC$-invariant.) 
\end{itemize} 
In each case, the value of the $\GC$-invariant vanishes 
on the constrained manifolds, 
since, even in the cases of the exceptional groups, 
the constraints $\del I = 0$ lead to $I=0$. 
This remarkable fact can be understood as follows: 
{\em Note that, in each case, 
the $\GC$-invariant $I$ is not invariant under a gauge group. 
Hence, it must vanish to be consistent with a gauge symmetry.}
We call this the ``consistency condition 
with a gauge symmetry''.\footnote{
By combining the result in Ref.~\cite{Ni2}, 
this condition can be understood as the condition that 
the manifold before gauging must be 
an open orbit, not a closed orbit. 
In Ref.~\cite{Ni2}, 
it was shown that an open orbit includes 
a compact and homogeneous manifold 
as a submanifold. 
Contrastingly, a closed orbit does not have 
such a submanifold.
}

If we forget the F-term constraints and impose only the 
D-term constraints,  
the manifolds become ${\bf C}P^{N-1}$ or $G_{2N,N}$. 
This means that all of the hermitian symmetric spaces 
are holomorphically embedded in 
${\bf C}P^{N-1}$ or $G_{N,M}$ by F-term constraints, 
as is shown in the last column of Table~2. 
Although some of the constraints are already known in 
the mathematical literature, 
the explicit forms of the constraints 
in the $E_6$ and $E_7$ cases 
are new results: 
$E_6/SO(10) \times U(1)$ is 
holomorphically embbedded in ${\bf C}P^{26}$ 
by $16$ quadratic homogeneous constraints, 
and $E_7/E_6 \times U(1)$ is embedded in ${\bf C}P^{55}$ 
by $28$ tripletic homogeneous constraints. 
The consistency condition with a gauge symmetry 
can be understood if we interpret the F-term constraints 
as the embedding conditions. 
Since $G_{N,M}$ can be embedded into ${\bf C}P^N$, 
all hermitian symmetric spaces are embedded in ${\bf C}P^N$. 
If we want to embed $M$ into ${\bf C}P^N$, 
the constraint must be homogeneous, 
when it is written in terms of homogeneous coordinates.\footnote{ 
The manifold, which can be embedded into ${\bf C}P^N$, 
is a (projective) algebraic variety 
and can be understood as a Hodge manifold.
}

In this paper, we have used the equation of motion for the 
vector auxiliary field. In the path integral formalism, 
this procedure corresponds to integrating over the 
vector field. In a separate paper~\cite{HN}, we show that 
the path integration can be performed exactly.

\medskip
Now we discuss possible generalizations of our results 
to wider class of K\"{a}hlerian $G/H$.
In this paper, we treated hermitian symmetric spaces, 
which are a special class of homogeneous K\"{a}hler manifolds. 
We confined ourselves to the gauge groups of $U(1)$ or $U(N)$. 
\begin{enumerate}
\item Even within this limitation, it is possible to 
generalize our construction to a wider class of homogeneous 
K\"{a}hler manifolds. Let us consider K\"{a}hler $G/H$, 
where $H$ has only one $U(1)$ factor, 
$H=H_{\rm ss}\times U(1)$, 
with $H_{\rm ss}$ being a semisimple subgroup of $H$. 
To be specific, let us generalize $SO(2N)/U(N)$. 
By generalizing $\Ph$ to a $2N \times M$ matrix ($N \geq M$),
transforming under $SO(2N) \times U(M)$ as
$\Ph \to g_{\rm L} \Phi {g_{\rm R}}^{-1}$, 
with the same superpotential (\ref{SOSp-superpot.}) 
(where $J$ is the same as in Eq.~(\ref{J})), 
we obtain 
\beq
  ({\bf R}^+)^{M^2}\times
 {SO(2N)_{\rm L} \times U(M)_{\rm R} \over 
      SO(2N-2M)_{\rm L}\times U(M)_{\rm V}}
    &\stackrel{U(M)_{\rm R}}{\longrightarrow}& 
  {SO(2N)_{\rm L} \over SO(2N-2M)_{\rm L}\times U(M)_{\rm L}}. 
 \nonumber
\eeq
This reduces to $SO(2N)/SO(2N-2) \times U(1)$ when $M=1$ 
and to $SO(2N)/U(N)$ when $N=M$.  
Similarly, $Sp(N)/U(N)$ can also be generalized. 
By generalizing $\Ph$ to a $2N \times M$ matrix ($N \geq M$), 
we obtain 
\beq
  ({\bf R}^+)^{M^2} \times
 {Sp(N)_{\rm L} \times U(M)_{\rm R} \over 
      Sp(N-M)_{\rm L} \times U(M)_{\rm V}}
    &\stackrel{U(M)_{\rm R}}{\longrightarrow}& 
  {Sp(N)_{\rm L} \over Sp(N-M)_{\rm L} \times U(M)_{\rm L}}. 
  \nonumber
\eeq
\item Now we consider generalization to 
the case of many $U(1)$ factors. 
Remember that the FI parameter $c$ becomes 
a decay constant, which represents the size of $G/H$, 
after integrating out the vector superfield.    
Then, we can consider 
there the be a one-to-one correspondence between
the decay constants and the FI-parameters.
Hence, to obtain $G/H$ with $H= H_{\rm ss} \times U(1)^n$ 
we must prepare $n$ FI-parameters. 
We thus consider a global symmetry, 
$G\pri = G\times G_1 \times \cdots G_n$, 
where each $G_i$ includes a $U(1)$ factor. 
If we gauge all $G_i$, 
the gauged K\"{a}hler potential has $n$ FI terms. 
After integrating out vector superfields, 
we obtain 
$G/H\pri \times G_1 \times \cdots G_n
=G/H_{\rm ss}\times U(1)^n$, 
where $H\pri$ is the remaining part after 
embedding all $G_i$ into $G$. 
Here we have put $H_{\rm ss}= H\pri \times {G_1}_{\rm ss} 
\times \cdots {G_n}_{\rm ss}$.
In the case of hermitian symmetric spaces, we have introduced 
an irreducible representation of $G$ as the basic field. 
It seems 
that we have to introduce more irreducible representations in 
these generalizations. 
Then we must impose orthogonality relations on these fields 
with D-term or F-term constraints. 
At the moment, we are unable to 
find consistent constraints in these cases.
\end{enumerate}

\section*{Acknowledgments}
The work of M.~N. is supported in part by 
JSPS Research Fellowships.

\begin{appendix}
\section{BKMU-IKK Construction of K\"{a}hler Potentials 
of Compact Homogeneous K\"{a}hler Manifolds}
Bando et al. (BKMU) gave the general method 
to construct the $G$ invariant K\"{a}hler potential 
of $G^{\bf C}/ \hat H$~\cite{BKMU}. 
However, there remained an ambiguity in the choice of 
the projection operators $\eta_i$ 
introduced below, Eq.~(\ref{proj.}).  
Itoh et al. (IKK) constructed these operators explicitly 
for the case that the target is compact, 
namely $G^{\bf C}/ \hat H \simeq G/H$~\cite{IKK}. 
Note that their method does not ensure that 
such models can be obtained from linear models.  
In this appendix, we review their method to 
compare with our method, 
which, on the other hand, has a linear origin. 

First of all, we need the projection matrices, 
which project a fundamental representation space 
onto a $\hat H$ invariant subspace~\cite{BKMU}. 
They satisfy the projection conditions 
\beq
 \eta\dagg = \eta, \hspace{1cm} 
 \eta \hat H \eta =\hat H \eta, \hspace{1cm} 
 \eta^2 = \eta.  \label{proj.}
\eeq
In an arbitrary K\"{a}hler $G/H$, 
the numbers of projection matrices 
is equals to the number of $U(1)$ factors in $H$. 
Since there is only one $U(1)$ factor 
in the hermitian symmetric cases, 
there is one projection matrix. 
In each case, it can be written as~\cite{IKK} 
\beq
 \eta = \pmatrix{ 1 &  \cr
                    & {\bf 0}} .
\eeq 
By using this, the K\"{a}hler potentials 
of compact K\"{a}hler manifolds 
can be written as~\cite{BKMU}
\beq
 K = c \log {\rm det}_{\eta} \xi\dagg\xi,
\eeq
where $\xi$ is a representative of 
the complex coset $\GC/\hat H$.
Since the form of $\xi$ can be calculated 
as Eqs.~(\ref{CPN-phi2}), (\ref{Gra-phi2}), (\ref{SO-phi}),
(\ref{SO,Sp-phi2}), (\ref{E6-phi}) and (\ref{E7-phi}), 
they give the same K\"{a}hler potential 
obtained from linear models in this paper.

\section{The Non-Compactness Theorem of Lerche and Shore}
The nonlinear sigma model, whose target manifold 
is compact and homogeneous, 
has a unique K\"{a}hler potential,  
as discussed in the last appendix~\cite{BKMU,IKK}. 
Although these models include neither a QNG boson nor 
an arbitrariness in the K\"{a}hler potential 
and they are mathematically beautiful, 
they cannot be obtained from any linear model, 
at least when there is no gauge symmetry:  
It was shown that there exists at least 
one QNG boson, and therefore 
the target must be non-compact and non-homogeneous. 
In this appendix, we review the theorem obtained 
by Lerche and Shore~\cite{LS} 
(see also Ref.~\cite{non-compact}). 
 
The fact that the model has a linear origin implies 
that the target manifold can be obtained from 
some F-term conditions (if there is no gauge symmetry). 
Since they are holomorphic equations, 
the invariance under the global symmetry $G$ 
enlarges to the complexification $\GC$, 
and the manifold becomes a $\GC$-orbit of 
the vacuum expectation value $v$.\footnote{
If there are not enough F-term constraints, 
the manifold may become larger than a $\GC$-orbit.
However, the proof is valid also in such cases, 
since they include at least one $\GC$-orbit.
} 
The pure-type multiplets require 
that the real broken generators 
are divided into complex unbroken and complex broken 
generators, $E^i$ and $\bar E_i (= (E^i)\dagg)$. 
Since $\bar E_i$ is broken, we obtain
\beq
 0 \neq |\bar E_i v|^2
 = v\dagg \left[ E^i, \bar E_i\right] v 
 = \alpha(i)^a v\dagg  H_a v,
\eeq
where $\alpha(i)^a$ is a root vector and 
$H_a$ is a Cartan generator. 
Therefore, at least one Cartan generator, 
$H_a$, must be broken. 
Since this is hermitian,  
there exists at least one mixed-type generator, 
and therefore at least one QNG boson.

\section{$SO(N)$ Algebra}
Since the basis of $SO(N)$ used in Subsec.~3.1 
is not in the standard form, 
here we give its relation to the ordinary basis. 
The $SO(N)$ generators in the ordinary basis are 
\beq
 {(T_{ij})^k}_l = 
 \1{i}(\delta^k_i \delta_{jl} - \delta^k_j \delta_{il}).
\eeq
In the basis, 
the vacuum expectation value 
satisfying $\vec{v}^{\,2} = 0$ can be written as 
\beq
 \vec{v} 
 = \pmatrix{- \2{\sqrt 2} \cr {\bf 0} \cr \1{\sqrt 2}}. 
\eeq
The real unbroken generators, 
at the center of the matrix, generate $SO(N-2)$.
The complex unbroken and broken generators are
\beq
E^i =
\left(
   \begin{array}{c|ccc|c}
             & \cdots & -\2{\sqrt 2} &\cdots&    \\ \hline
      \vdots &        &             &      & \vdots   \\
 \2{\sqrt 2} &        &{\bf 0}_{N-2}&      & -\1{\sqrt 2}\\ 
      \vdots &        &             &      & \vdots \\ \hline
             & \cdots & \1{\sqrt 2} &\cdots&           
   \end{array}
   \right), \hspace{0.5cm}
\bar E_i =
\left(
   \begin{array}{c|ccc|c}
            & \cdots & -\2{\sqrt 2}&\cdots&    \\ \hline
     \vdots &        &             &      & \vdots   \\
 \2{\sqrt 2}&        &{\bf 0}_{N-2}&      & \1{\sqrt 2}\\ 
     \vdots &        &             &      & \vdots \\ \hline
            & \cdots & -\1{\sqrt 2}&\cdots&           
   \end{array}
   \right),
\eeq
where $i=1,\cdots,N-2$ and 
only the $i$-th components are nonzero. 
The broken $U(1)$ generator is 
\beq
T = 
\left(
   \begin{array}{c|c|c}
         &               &-i \\ \hline
         & {\bf 0}_{N-1} &   \\ \hline
        i&               &  
   \end{array}
   \right) .
\eeq
This generator will become unbroken 
after gauging $U(1)_{\rm D}$. 
Here, we change the basis by a unitary transformation with  
\beq
 U = \left(
     \begin{array}{c|c|c}
      \2{\sqrt 2}&               &\1{\sqrt 2}  \\ \hline
                & {\bf 1}_{N-1} &  \\ \hline
     -\2{\sqrt 2}&               &\1{\sqrt 2}  
     \end{array}
     \right) .
\eeq
Since $U$ is a unitary matrix ($U\dagg U = U U\dagg =1$), 
$\vec{\phi}^{\,\dagger}\vec{\phi}$ is invariant, 
and then $\log (\vec{\phi}^{\,\dagger}\vec{\phi})$ 
also is invariant. 
By the unitary transformation, 
the vacuum expectation value is transformed 
to the standard form, 
\beq
 U \vec v = \pmatrix{1 \cr {\bf 0}}.
\eeq
The $SO(N-2)$ generators are not transformed,  
and the other generators are transformed as 
\beq
&&U E^i U\dagg =
\left(
   \begin{array}{c|ccc|c}
         & \cdots & 1           &\cdots&    \\ \hline
  \vdots &        &             &      & \vdots   \\
     0   &        &{\bf 0}_{N-2}&      & -1\\ 
  \vdots &        &             &      & \vdots \\ \hline
         & \cdots &   0         &\cdots&           
   \end{array}
   \right), \;\;\;
U \bar E_i U\dagg = 
\left(
   \begin{array}{c|ccc|c}
         & \cdots &  0          &\cdots&    \\ \hline
  \vdots &        &             &      & \vdots   \\
       1 &        &{\bf 0}_{N-2}&      & 0\\ 
  \vdots &        &             &      & \vdots \\ \hline
         & \cdots & -1          &\cdots&           
   \end{array}
   \right), \non
&&U T U\dagg= 
\left(
   \begin{array}{c|c|c}
        1&               &   \\ \hline
         & {\bf 0}_{N-1} &   \\ \hline
         &               &-1  
   \end{array}
   \right) .
\eeq
We thus obtain the transformation law (\ref{SO-tr.law}) 
used in Subsec.~3.1.
Moreover, the second rank invariant tensor 
is transformed as  
$\delta_{ij} \to (U \delta U^T)_{ij} = J_{ij}$, 
where $J$ is defined in Eq,~(\ref{J-SO}).

\section{$E_6$ Algebra}
In this appendix, 
we construct the $E_6$ algebra 
by referring to Refs.~\cite{IKK2,KuSa}. 

\subsection{Construction of $E_6$ algebra}
Since an adjoint representation is 
decomposed as ${\bf 78} = ({\bf 45},0) \oplus ({\bf 1},0)
\oplus({\bf 16},1)\oplus ({\bf \bar{16}},-1)$~\cite{Sl}, 
we construct the $E_6$ algebra as 
${\cal E}_6 = {\cal SO}(10) \oplus {\cal U}(1) 
\oplus {\bf 16} \oplus {\bf \bar{16}}$: 
We prepare the $SO(10)$ generator $T_{AB}$, 
the $U(1)$ generator $T$, 
${\bf 16}$ as $E_{\alpha}$, and 
${\bf \bar{16}}$ as $\bar{E}^{\alpha} = (E_{\alpha})\dagg$. 
Then their commutation relations can be calculated 
as follows~\cite{IKK2,KuSa}: 
\beq
 &&[T_{AB},T_{CD}] 
 = -i(\delta_{BC}T_{AD} + \delta_{AD}T_{BC}
    - \delta_{AC}T_{BD} - \delta_{BD}T_{AC}), 
   \;\; [T, T_{AB}] = 0, \non
 &&[T_{AB},E_{\alpha}] 
     = -{({\sigma_{AB}})_{\alpha}}^{\beta}E_{\beta}, 
   \hspace{0.4cm} 
   [T_{AB},\bar{E}^{\alpha}] 
     = {({\sigma^*_{AB}})^{\alpha}}_{\beta} \bar{E}^{\beta}, \non
 &&[T, E_{\alpha}] = {\sqrt 3\over 2} E_{\alpha}, \hspace{2cm} 
   [T, \bar{E}_{\alpha}] = -{\sqrt 3\over 2} \bar{E}^{\alpha},\non
 &&[E_{\alpha},E_{\beta}] 
  = [\bar{E}^{\alpha},\bar{E}^{\beta} ] = 0,  \hspace{0.7cm}
   [E_{\alpha} ,\bar{E}^{\beta} ]
 = -\1{2}{({\sigma_{AB}})_{\alpha}}^{\beta} T_{AB} 
   + {\sqrt 3 \over 2} {\delta_{\alpha}}^{\beta} T.
\eeq
The $U(1)$ charge of $E_{\alpha}$ is determined by 
the difference between $U(1)$ charges 
of $x$ and $y$ or $y$ and $z$ 
in Eq.~(\ref{E6-tr.}): 
${2 \over \sqrt{3}} - {1\over 2\sqrt{3}} 
= {1\over 2\sqrt{3}} - (-\1{\sqrt 3})= {\sqrt 3 \over 2}$.
The second coefficient of the last equation has 
the same value as the $U(1)$ charge of $E_{\alpha}$, from 
the anti-symmetric property of the structure constants. 
The relative weight of the first and the second terms  
is determined by using the Jacobi identity, 
$[\bar{E},[E,E]] + {\rm (cyclic)} = 0$, 
and the nontrivial identity for 
the spinor generators~\cite{IKK2,KuSa},
\beq
 \Sigma_{AB}{({\sigma_{AB}})_{\alpha}}^{[\beta}
 {({\sigma_{AB}})_{\gamma}}^{\delta]}
  = {3\over 2} {\delta_{\alpha}}^{[\beta}{\delta_{\gamma}}^{\delta]}.
\eeq

\subsection{Invariant tensor of $E_6$}
From the tensor product~\cite{Sl} 
${\bf 27} \otimes {\bf 27} = {\bf \bar {27}}_{\rm s} \oplus \cdots$, 
we know there exists a rank-$3$ symmetric 
tensor invariant under $E_6$. 
The components of $\Gamma_{ijk}$ are~\cite{KuSa}
\beq
 \Gamma_{ijk} = \cases{
  \Gamma_{0AB} = \delta_{AB}, \cr
  \Gamma_{A\alpha\beta} 
     = \1{\sqrt 2}(C{\sigma_A}\dagg)^{\alpha\beta},  \cr
  \mbox{otherwise }0 .}    \label{E6inv.tensor}
\eeq
These components can be calculated as follows. 
First, construct the $SO(10)\times U(1)$ invariant 
of order three: 
\beq
 I_3 = A x z^2 + \1{\sqrt 2} 
   z^A y_{\alpha} (C{\sig_A}\dagg)^{\alpha\beta} y_{\beta} .
\eeq
By the requirement of the invariance of $E$ or $\bar E$, 
we can conclude $A=1$. 
(Here we have used the identity (\ref{spinor-id.}).)  
The components (\ref{E6inv.tensor}) can 
be read from this invariant.

It is known that there is an identity~\cite{KV}~\footnote{
In the calculation of $\Gamma_{ijA}\Gamma^{ijB} = 10 \delta_{AB}$, 
we have used the identity
$2^{-4} \tr(C{\sigma_A}\dagg\sigma_BC)=\delta_{AB}$~\cite{KuSa}.}
\beq
 \Gamma_{ijk}\Gamma^{ijl} = 10 \delta_k^l .\label{E6-id.}
\eeq
Under the normalization in Eq.~(\ref{E6-id.}), 
there is the Springer relation~\cite{KV} 
\beq
 \Gamma_{ijk}(\Gamma^{jl\{m}\Gamma^{no\}k}) 
 = {\delta_i}^{\{l} \Gamma^{mno\}} ,\label{Springer}
\eeq
where we have used the notation 
$A^{\{ij\cdots\}}=A^{ij\cdots} + A^{ji\cdots} + \cdots$.
These identities are used many times 
in the analysis of the $E_7$ algebra.

\section{$E_7$ Algebra}
In this appendix, 
we construct the $E_7$ algebra in the same way 
as in the last appendix.

\subsection{Construction of $E_7$ algebra}
The decomposition of the adjoint representation of $E_7$ 
under the maximal subgroup $E_6 \times U(1)$ is 
${\bf 133} = ({\bf 78},0) \oplus ({\bf 1},0)
\oplus ({\bf 27},1) \oplus ({\bf \bar{27}},-1)$, 
where the second components are the $U(1)$ charges~\cite{Sl}.
Hence, we can construct the $E_7$ algebra 
by adding generators $E^i$ and $\bar E_i (= (E^i)\dagg)$ 
($i=1,\cdots,27$), which belong to 
the $E_6$ fundamental and anti-fundamental 
representations, respectively, 
to the $E_6 \times U(1)$ algebra, 
$T_A$ ($A=1,\cdots,78$) and $T$: 
${\cal E}_7 = {\cal E}_6 \oplus {\cal U}(1) 
\oplus {\bf 27} \oplus {\bf \bar{27}}$.
In the same manner as we constructed 
the $E_6$ algebra in the last appendix, 
their commutation relations are obtained as follows: 
\beq
 &&[T_A,T_B] = i {f_{AB}}^C T_C, \hspace{1.1cm} 
   [T, T_A] = 0, \non
 &&[T_A,E^i] = {\rho(T_A)^i}_j E^j, \hspace{1cm} 
   [T_A,\bar{E}_i] = -{{\rho(T_A)^T}_i}^j \bar{E}_j, \non
 &&[T, E^i] =  \sqrt{2\over 3} E^i, \hspace{1.9cm} 
   [T, \bar{E}_i] = - \sqrt{2\over 3}\bar{E}_i, \non
 &&[E^i,E^j] = [\bar{E}_i,\bar{E}_j ] = 0,  \hspace{0.7cm}
   [E^i,\bar{E}_j] 
 = {\rho(T_A)^i}_j T_A + \sqrt{2 \over 3} {\delta^i}_j T .
\eeq
Here $\rho(T_A)$ is a fundamental representation matrix, 
and the ${f_{AB}}^C$ are structure constants of $E_6$, 
whose explicit forms were obtained in the last section. 
The $U(1)$ charge of $E^i$ is determined from 
the difference of $x$ and $y^i$, etc., 
in Eq.~(\ref{E7_action}), and $\bar E_i$ is its conjugate. 
In the last equation, 
the coefficient of the second term coincides with 
the $U(1)$ charge of $E^i$ due to the anti-symmetricity 
of the structure constants of $E_7$. 
The first term is determined by 
the Jacobi identity $[\bar E,[E,E]] + ({\rm cyclic}) = 0$ and 
the nontrivial identity for 
the $E_6$ fundamental representation~\cite{Cv},
\beq
 \Sigma_A \; {\rho(T_A)^{[i}}_j {\rho(T_A)^{k]}}_l
  = -{2\over 3} {\delta^{[i}}_j {\delta^{k]}}_l  .
\eeq
This is satisfied when $\tr \rho(T_A)^2 = 6$.

\subsection{Invariant tensors of $E_7$}
From the tensor product of 
fundamental representations~\cite{Sl}, 
${\bf 56} \otimes {\bf 56} = {\bf 1}_{\rm a}\oplus \cdots$ and
${\bf 56} \otimes {\bf 56}\otimes {\bf 56}\otimes {\bf 56} 
= {\bf 1}_{\rm s}\oplus \cdots$, 
there exist the rank-2 anti-symmetric tensor $f_{\alpha\beta}$ 
and the rank-4 symmetric tensor $d_{\alpha\beta\gamma\delta}$ 
as $E_7$ invariant tensors. 
To find their components, 
we construct a linear combination of 
$E_6\times U(1)$ invariants of quartic order 
and require invariance under $E$ or $\bar E$, 
as in the last appendix. 
The result is 
\beq
 I_4 &=& d_{\alpha\beta\gamma\delta} 
     \phi^{\alpha} \phi^{\beta} \phi^{\gamma} \phi^{\delta}\non
  &=& -\1{2} (x w - y^i z_i)^2 
   -\1{3}w \Gamma_{ijk} y^iy^jy^k -\1{3}x \Gamma^{ijk} z_iz_jz_k\non
   &&+\1{2} \Gamma^{ijk}\Gamma_{ilm} z_j z_k y^l y^m  . 
\eeq
Here, $I_4$ is invariant due to the Springer relation 
for the $E_6$ invariant tensor, Eq.~(\ref{Springer}). 
The components can be read from this invariant. 
Since we do not use the anti-symmetric tensor 
$f_{\alpha\beta}$, we do not construct it here.

\end{appendix}


\end{document}